\documentclass[lettersize,journal]{IEEEtran}
\usepackage{amsmath,amsfonts}
\usepackage{array}
\usepackage{textcomp}
\usepackage{stfloats}
\usepackage{url}
\usepackage{verbatim}
\usepackage{graphicx}
\usepackage{cite}
\usepackage{amsthm,amsbsy,amssymb,amsfonts,booktabs,subfigure,multirow,wasysym,algpseudocode,enumitem,threeparttable,hyperref}
\usepackage[ruled,vlined,linesnumbered]{algorithm2e}
\usepackage{soul,color,xcolor,colortbl}
\soulregister\cite7
\soulregister\citep7
\soulregister\citet7
\soulregister\ref7
\soulregister\pageref7
\soulregister\em7

\definecolor{myColor}{rgb}{0,0,0}        
\makeatletter
\newcommand*{\revise}{\@ifnextchar\bgroup{\revise@}{\color{myColor}}}
\newcommand*{\revise@}[1]{{\textcolor{myColor}{#1}}}
\makeatother

\begin{document}

\title{From Static to Adaptive Defense: Federated Multi-Agent Deep Reinforcement Learning-Driven Moving Target Defense Against DoS Attacks in UAV Swarm Networks}

\author{Yuyang~Zhou,~\IEEEmembership{Member,~IEEE},~Guang~Cheng,~\IEEEmembership{Member,~IEEE},~Kang~Du,~\IEEEmembership{Student Member,~IEEE},\\~Zihan~Chen,~\IEEEmembership{Member,~IEEE},~Tian~Qin,~and~Yuyu~Zhao,~\IEEEmembership{Member,~IEEE}
 \thanks{Yuyang Zhou, Guang Cheng, Kang Du, Zihan Chen, Tian Qin, and Yuyu Zhao are with the School of Cyber Science and Engineering, Southeast University, Purple Mountain Laboratories, and Jiangsu Province Engineering Research Center of Security for Ubiquitous Network, Nanjing 211189, China. E-mail: \{yyzhou, chengguang, dukang, zhchen\_seu, 230208959, yyzhao\}@seu.edu.cn.}
 \thanks{Guang Cheng is the corresponding author.}}

\markboth{Journal of \LaTeX\ Class Files,~Vol.~14, No.~8, August~2021}%
{Shell \MakeLowercase{\textit{\emph{et al.}}}: A Sample Article Using IEEEtran.cls for IEEE Journals}


\maketitle

\begin{abstract}
  The proliferation of unmanned aerial vehicles (UAVs) has enabled a wide range of mission-critical applications and is becoming a cornerstone of low-altitude networks, supporting smart cities, emergency response, and more. However, the open wireless environment, dynamic topology, and resource constraints of UAVs expose low-altitude networks to severe Denial-of-Service (DoS) threats, undermining their reliability and security. Traditional defense approaches, which rely on fixed configurations or centralized decision-making, cannot effectively respond to the rapidly changing conditions in UAV swarm environments. To address these challenges, we propose a novel federated multi-agent deep reinforcement learning (FMADRL)-driven moving target defense (MTD) framework for proactive DoS mitigation in low-altitude networks. Specifically, we design lightweight and coordinated MTD mechanisms, including leader switching, route mutation, and frequency hopping, to disrupt attacker efforts and enhance network resilience. The defense problem is formulated as a multi-agent partially observable Markov decision process (POMDP), capturing the uncertain nature of UAV swarms under attack. Each UAV is equipped with a policy agent that autonomously selects MTD actions based on partial observations and local experiences. By employing a policy gradient-based FMADRL algorithm, UAVs collaboratively optimize their policies via reward-weighted aggregation, enabling distributed learning without sharing raw data and thus reducing communication overhead. Extensive simulations demonstrate that our approach significantly outperforms state-of-the-art baselines, achieving up to a 34.6\% improvement in attack mitigation rate, a reduction in average recovery time of up to 94.6\%, and decreases in energy consumption and defense cost by as much as 29.3\% and 98.3\%, respectively, under various DoS attack strategies. These results highlight the potential of intelligent, distributed defense mechanisms to protect low-altitude networks, paving the way for reliable and scalable low-altitude economy.
\end{abstract}

\begin{IEEEkeywords}
Low-altitude networks, unmanned aerial vehicles, denial-of-service attacks, moving target defense, federated multi-agent deep reinforcement learning.
\end{IEEEkeywords}

\section{Introduction}
\IEEEPARstart{T}{he} rapid development of unmanned aerial vehicle (UAV) technology~\cite{zuo2022unmanned} has enabled a wide range of applications, including environmental monitoring, disaster response, precision agriculture, logistics, aerial photography, and intelligent surveillance~\cite{xie2025disrupting}. By leveraging the collaborative capabilities of multiple UAVs~\cite{sun2025aerial}, the UAV swarm~\cite{javed2024state,cao2024computational} can achieve enhanced coverage, resilience, and real-time data processing~\cite{wang2024generative}, making them indispensable in both civilian and industrial domains. It is expected to play an increasingly important role in smart cities, emergency management, and next-generation communication infrastructures, forming the backbone of low-altitude networks.

Nevertheless, the widespread adoption of UAV swarms also brings new security challenges~\cite{tsao2022survey,wang2025safeguarding} to low-altitude networks. Due to their reliance on open wireless links, limited energy and processing capabilities, UAV networks are particularly vulnerable to \emph{Denial-of-Service} (DoS) attacks~\cite{tang2024secure, yang2025resilient}. For example, UAVs often operate without robust authentication or traffic filtering mechanisms. Attackers can easily launch attacks by overwhelming low-altitude communication channels or computational resources, leading to service disruptions or UAV disconnection from the swarm~\cite{shi2025neural}. In mission-critical scenarios, even a brief loss of connectivity or control can have catastrophic consequences, including the failure of time-sensitive missions or the crash of drones.

In addition, UAV swarms operate in highly dynamic, resource-constrained, and often uncertain environments. The mobility of UAVs, the need for low-latency communication, and the distributed nature of control introduce unique challenges for both attack detection and defense\cite{wang2025generative, wang2023survey}. Traditional security mechanisms, such as firewalls~\cite{ucctu2021suggested}, intrusion detection systems (IDSs)~\cite{zhong2024survey}, and traffic redirection, typically rely on prior knowledge of attack characteristics and are often designed for centralized infrastructures, making them ill-suited for low-altitude networks~\cite{zhao2024two}. Moreover, the static nature of these defenses results in several shortcomings: (i) limited adaptability to evolving and sophisticated attack patterns, (ii) increased false positive rates due to lack of contextual awareness, and (iii) reactive mitigation that only occurs after attacks have already caused damage. Therefore, ensuring the resilience of low-altitude networks against DoS attacks by more proactive and adaptive defense strategies is a critical challenge.

To address this, \emph{Moving Target Defense} (MTD)~\cite{tan2023survey, zhang2025moving} has emerged as a promising security paradigm that aims to increase the complexity and cost for attackers by continuously or periodically changing the system's attack surface~\cite{zhou2025resource}. Instead of relying on static configurations, MTD introduces dynamic and unpredictable changes, such as IP randomization and port hopping, so that attackers cannot easily gather information or exploit vulnerabilities. This approach fundamentally shifts the traditional defender-attacker asymmetry, making it significantly harder for adversaries to plan and execute successful attacks.

However, applying traditional MTD to UAV scenarios presents three critical challenges. (i) First, many MTD mechanisms consume significant resources, making them unsuitable for resource-constrained UAV platforms. (ii) Second, most MTD decision-making frameworks are designed to collect global information and coordinate defense actions in a centralized manner, which cannot operate efficiently and reliably within the unique constraints of low-altitude networks. (iii) Third, sophisticated attackers can reconstruct attack target states by tracking and learning the mutation patterns, enabling them to persistently target UAVs despite the use of MTD. These limitations highlight the urgent need for distributed and intelligent MTD framework that can adaptively balance security and performance in UAV scenarios.

\begin{figure}[t]
  \centering
  \includegraphics[width=0.95\columnwidth]{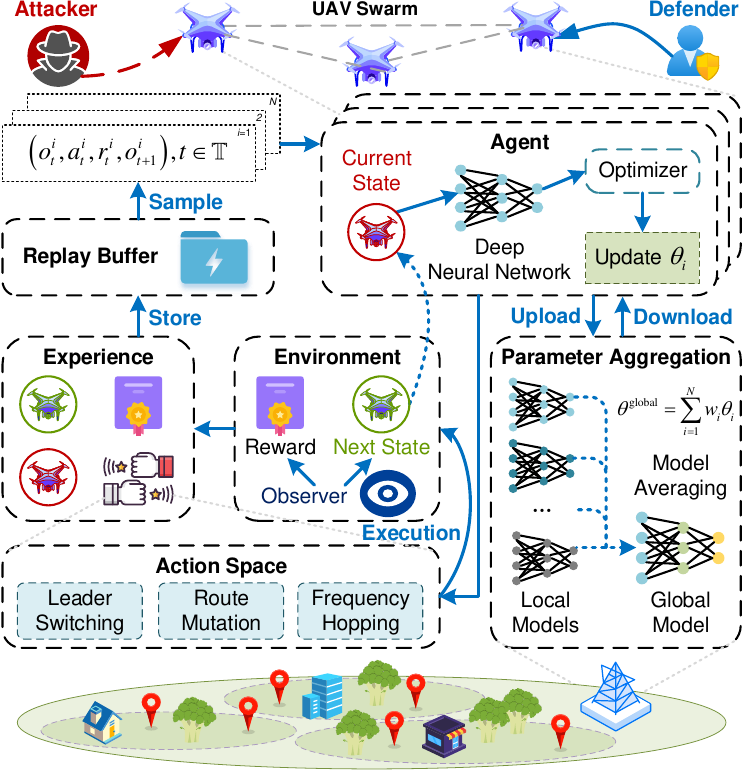}
  \caption{Bird's-eye view of the proposed DoS mitigation approach.}
  \label{Scenario}
\end{figure}

In this paper, we propose a novel \emph{federated multi-agent deep reinforcement learning} (FMADRL)-driven MTD framework tailored for UAV swarm networks facing DoS attacks, where the bird's-eye view of the proposed method has been illustrated in Fig.~\ref{Scenario}. In our framework, we design three lightweight MTD actions, including (i) \emph{Leader Switching}, (ii) \emph{Route Mutation}, and (iii) \emph{Frequency Hopping}, based on partial observations. The first mechanism allows the UAV swarm to promptly reassign the leader role among eligible UAVs when the current leader is persistently targeted, while the second mechanism dynamically reconfigures communication paths by selecting alternative relay UAVs, ensuring that critical messages can bypass compromised links and reach their destinations. Furthermore, frequency hopping periodically changes the communication frequency channels used by the swarm, significantly increasing the uncertainty for attackers and disrupting their ability to sustain effective attacks over time. To effectively mitigate adaptive DoS attacks while minimizing resource consumption and maintaining mission continuity, we first formulate the defense problem as a multi-agent partially observable Markov decision process (POMDP) and develop a policy gradient-based FMADRL (PG-FMADRL) method, where each UAV is treated as an independent agent that can learn and adapt its defense strategies based on local observations and experiences. Then, each agent periodically shares only its model parameters with a central aggregator using a federated learning scheme. This collaborative approach enables the UAV swarm to adaptively coordinate distributed defense strategies in real time without sharing raw data, thus respecting the stringent resource and latency constraints of low-altitude networks. The main contributions of this paper are summarized as follows:

\begin{itemize}[leftmargin=*]
  \item \textbf{Proactive and collaborative DoS defense framework.} We propose a novel framework for DoS mitigation in UAV swarm networks that enables self-adaptive defense among distributed UAV nodes. Our proposed approach eliminates the need for attack detection and advances security by establishing a proactive and collaborative paradigm.
  \item \textbf{Lightweight and adaptive MTD mechanisms.} Within the proposed framework, we consider the effects of heavy consumption and high delay issues that arise in existing MTD solutions, and thus design three lightweight, flexible, and adaptive MTD mechanisms specifically optimized for dynamic and resource-constrained UAV swarm environments.
  \item \textbf{Federated and intelligent defense decision-making.} We formulate the distributed defense problem as a multi-agent POMDP, capturing the uncertainty and limited observability in UAV swarm networks under attack. Based on this formulation, we develop a PG-FMADRL method that enables UAVs to collaboratively learn and optimize defense policies.
  \item \textbf{Comprehensive evaluation of system performance.}
 Through extensive simulations, we demonstrate that our method outperforms the state-of-the-art (SOTA) schemes in terms of attack mitigation rate, while imposing less recovery time and overhead. The source code is available at \url{https://github.com/SEU-ProactiveSecurity-Group/PG-FMADRL}.
\end{itemize}

The remainder of this paper is organized as follows. We first review related work on DoS mitigation and MTD techniques in Section~\ref{section2}. Next, Section~\ref{section3} introduces the system architecture and problem formulation. The details of the proposed FMADRL-driven MTD framework are presented in Section~\ref{section4}, followed by simulation results and performance analysis in Section~\ref{section5}. Finally, Section~\ref{section6} concludes the paper and discusses potential future research directions.

\section{Related Work}\label{section2}

This section reviews relevant literature on DoS attack mitigation in UAV networks and MTD-based defense mechanisms. We identify limitations in existing approaches and highlight the research gap our work addresses.

\subsection{DoS Attacks Detection and Mitigation in UAV Networks}
\revise{Research on securing UAV networks against DoS attacks has largely focused on two areas, including detection and mitigation. Detection methods often employ machine learning models like convolutional neural network (CNN)~\cite{fu2023machine}, long short-term memory (LSTM)~\cite{hassler2024cyber} or conditional generative adversarial network (CGAN)~\cite{he2023cgan} to identify malicious traffic. While achieving high accuracy, these approaches rely heavily on large, labeled datasets, making them less effective against novel or zero-day attacks and difficult to deploy in real-world scenarios where such data is scarce.}

\revise{In addition to detection, mitigation strategies often involve adaptive control or game-theoretic approaches to enhance swarm resilience~\cite{wu2024zero}. For example, Tan \emph{et al.}~\cite{tan2025strategy} proposed a time-delay evolutionary game method for honeypoint defense, ensuring optimal strategy stability and selects the best defense at the 35th time unit on IEEE 22-Bus. These methods typically focus on tolerating or isolating disruptions rather than proactively preventing them. A key limitation of such reactive strategies is that they often act only after an attack has already caused damage, which may be insufficient against sophisticated adversaries who can adapt their tactics to circumvent static or predictable defenses.}

Compared with the above methods, our approach offers two significant advantages. First, our method does not rely on prior knowledge, labeled data, or attack signatures, making it more robust to novel and adaptive threats. Second, by orchestrating multiple lightweight MTD mechanisms, our framework can rapidly respond to attacks in real time, minimizing service disruption and resource consumption. These features collectively distinguish our work from prior studies and address key challenges in securing low-altitude networks.

\subsection{MTD-based Solutions for DoS/DDoS Attacks}
\revise{MTD has emerged as a proactive paradigm to counter DoS or DDoS attacks by dynamically altering the system's attack surface~\cite{wu2022secure,sanoussi2023itc}. In traditional scenarios, MTD often leverages software-defined networking (SDN) or virtualization to shuffle IP addresses or migrate services~\cite{ribeiro2023detecting, zhang2023moving}. However, these solutions present a significant limitation: their high computational and communication overhead makes them unsuitable for resource-constrained environments like low-altitude networks.}

\revise{To address this, some research has explored lightweight MTD methods for Internet of Things (IoT) and edge environments, using deep reinforcement learning (DRL) to optimize configuration changes~\cite{zhou2022toward,zhou2025resource}. For instance, Zhang \emph{et al.}~\cite{zhang2023mitigate} designed configuration mutation mechanisms against DDoS attacks, where the communication ranges and capacities of roadside units (RSUs) are dynamically adjusted. While more efficient, these methods face another key limitation. They are not directly applicable to the highly distributed and mobile nature of UAV swarms, where centralized decision-making creates bottlenecks and single points of failure.}

\revise{In contrast to these prior works, our framework is the first to integrate lightweight, coordinated MTD mechanisms with a federated multi-agent reinforcement learning approach. This design specifically addresses the current limitations of existing MTD solutions by enabling proactive, distributed, and intelligent defense tailored for the unique challenges of low-altitude networks.}

\section{System Model and Problem Formulation}\label{section3}
This section presents the system architecture and threat model that form the foundation of our work. We first introduce the UAV swarm model, then define the DoS attack types and attacker strategies, followed by our proposed MTD mechanisms. Finally, we formulate the defense problem as a multi-objective optimization problem and analyze its characteristics.

\subsection{System Model}
We consider a multi-UAV swarm system composed of a ground control station (GCS) and $N$ UAVs, denoted as $\mathcal{V} = \{0, 1, ..., N\}$, where node $0$ is the GCS and nodes $1$ to $N$ are UAVs. The system operates in a three-dimensional space and is tasked with persistent patrol and robust formation maintenance.

At any given time, one UAV is designated as the leader, responsible for receiving high-level commands from the GCS and relaying them to the rest of the swarm. The remaining UAVs act as followers, adjusting their positions based on local observations and received commands.

To understand communication possibilities within the swarm, we represent the network as a dynamic undirected graph $\mathcal{G}(t) = (\mathcal{V}, \mathcal{E}(t))$. An edge $(i, j) \in \mathcal{E}(t)$ exists if and only if UAVs $i$ and $j$ are within a communication range $R_c$, operate on the same frequency channel $f$ at time $t$. Formally,
\begin{equation}
  \label{eq1}
 (i, j) \in \mathcal{E}(t) \iff |\mathbf{p}_i(t) - \mathbf{p}_j(t)| \leq R_c,\quad f_i(t) = f_j(t)
\end{equation}
where $\mathbf{p}_i(t)$ denotes the position of UAV $i$ at time $t$.

\revise{The UAV swarm is required to maintain a circular formation of radius $r$ centered at $\mathbf{c} = (c_x, c_y, h)$. The ideal position for UAV $i$ at time $t$ is given by}
\begin{equation}
  \label{eq2}
  \revise{\mathbf{p}_i^*(t) = \mathbf{c} + r \begin{bmatrix}
    \cos \theta_i(t) \\
    \sin \theta_i(t) \\
    0
  \end{bmatrix}}
\end{equation}
where $\theta_i(t)$ is the desired angular position of UAV $i$. The formation rotates at a constant angular speed $\omega$, so that
\begin{equation}
  \label{eq3}
  \theta_i(t) = \theta_i(0) + \omega t
\end{equation}
where $\theta_i(0)$ is the initial angular position of UAV $i$. The patrol (linear) speed of each UAV along the circle is given by $v_{\text{pat}} = r \omega$. To ensure collision avoidance, the UAVs are required to maintain a minimum distance $d_{\min}$ from each other at all times such that
\begin{equation}
  \label{eq4}
  |\mathbf{p}_i(t) - \mathbf{p}_j(t)| \geq d_{\min}, \quad \forall i, j \in \mathcal{V} \setminus \{0\}
\end{equation}
where $d_{\min}$ is the minimum separation distance between UAVs. The maximum speed of each UAV is denoted as $v_{\max}$, which is the upper limit on the flying speed of the UAVs.

At each time step, each UAV updates its position and heading to minimize the deviation from its ideal position $\mathbf{p}_i^*(t)$ with its velocity $v_i(t)$. If the deviation exceeds a threshold $\delta$, i.e., $|\mathbf{p}_i(t) - \mathbf{p}_i^*(t)| > \delta$, the UAV will accelerate and move toward its ideal position at maximum speed $v_{\max}$ until the deviation falls below the threshold. Otherwise, it continues patrolling at the nominal speed $v_{\text{pat}}$.

\subsection{Threat Model}
In this work, we consider a UAV swarm network operating in an adversarial environment and the GCS is trusted, where external attackers aim to disrupt the swarm's communication and coordinated behavior. To comprehensively analyze the system's resilience and develop effective defense, we define the threat model in terms of the attack types and their strategies.

\subsubsection{Attack Types}
The attacker can launch two types of DoS attacks against the UAV swarm communication system according to attack targets as follows.
\begin{itemize}
\item \textbf{Node Attack}: The attacker targets a specific UAV $i$ and occupies its computation resources by sending a large amount of malicious requests through the current frequency~\cite{wang2021security}. When the UAV $i$ is under node attack at time $t$, it can be formally described as $\phi^\mathcal{N}_i(t) = 1$. 

\item \textbf{Link Attack}: The attacker targets a specific communication link $(i, j)$ between two UAVs or the GCS and the leader UAV, jamming the wireless channel by transmitting interference signals~\cite{gong2023resilient}. When the link $(i, j)$ is under DoS attack at time $t$, we can describe the communication between UAVs $i$ and $j$ as $\phi^\mathcal{L}_{ij}(t) = 1$.
\end{itemize}

\subsubsection{Attacker Strategies}\label{section3.2.2}
To capture a wide range of realistic adversarial behaviors, we consider three types of attacker strategies, such that
\begin{itemize}
  \item \textbf{Fixed Attacker}: The attacker selects a specific UAV or communication link as the attack target at the beginning of the DoS attack and persistently attacks it throughout the whole attack-defense interaction~\cite{yu2023reinforcement}. For example, if the fixed attacker selects a UAV node $i$ as the target with initialized attack frequency $f_{\text{atk}}$, then, this attack remains active for the entire interaction and can be described as $\phi^\mathcal{N}_i(t) = 1$ and $f_{\text{atk}}(t) = f_i(0)$ for all time steps $t \in \mathbb{T} = \{1, 2, \ldots, T\}$, where $T$ denotes the time horizon.
  \item \textbf{Random Attacker}: At each attack opportunity, the attacker randomly selects a UAV node or communication link from the set of available targets~\cite{boualouache2022federated}. The attacker launches an attack on the chosen target for a fixed attack duration $\tau_{\text{atk}}$. After each attack, the attacker needs to wait for a reconnaissance period $\tau_{\text{recon}}$ before initiating the next attack. During this reconnaissance period, the attacker is assumed to be passively listening or actively scanning to re-acquire a target, by identifying the swarm's current communication frequency if a frequency hop has occurred. For example, if the random attacker selects a UAV $i$ as the target at time step $0$, then the attack can be described as $\phi^\mathcal{N}_i(t) = 1$ and $f_{\text{atk}}(t) = f_i(0)$ for $t \in [0, \tau_{\text{atk}} + \tau_{\text{recon}})$, and $\phi^\mathcal{N}_j(t) = 0$ for all other UAVs $j \neq i$. Thus, the next attack can only start after initial attack ends and the attack frequency is updated to match the current frequency of the another selected UAV $j$, i.e., $f_{\text{atk}}(\tau_{\text{atk}} + \tau_{\text{recon}}) = f_j(\tau_{\text{atk}} + \tau_{\text{recon}})$.
  \item \textbf{Greedy Attacker}: The adversary dynamically observes the network state, enabling adaptively selecting the target that is expected to cause the maximum disruption to the swarm~\cite{huang2022strategic}. \revise{For example, the greedy attacker can select the current leader UAV $\mathbb{L}^\mathcal{N}(t)$ or the core link $\mathbb{L}^\mathcal{L}(t)$ of the swarm as the target at time step $t$.} For the sake of not weakening the greedy attacker's ability, we assume that he/she may be aware of the existence of defense mechanisms (e.g., MTD), but not their specific deployment or timing. Therefore, the attacker can re-evaluate and update his/her target at time step $t'$ to maximize attack effectiveness, but attack actions are subject to duration and reconnaissance constraints that are similar to those of the random attacker, i.e., $t' = t + \tau_{\text{atk}} + \tau_{\text{recon}}$. This reconnaissance period represents the time required for the adversary to gather new intelligence, such as identifying the new leader after a switch or discovering the new communication channel after a frequency hop, before launching the next targeted attack.
\end{itemize}

The effectiveness of an attack depends on the frequency alignment between the attacker and the victim, as well as the lasting time of attack duration and reconnaissance. For the $n$th node attack, it can be formally described as 
\begin{equation}
  \label{eq5}
E^\mathcal{N}_i(t) = 
\begin{cases}
1, & f_i(t) = f_{\text{atk}}(t) \land t \in [(n\!-\!1)\tau_{\text{eff}}, n\tau_{\text{eff}}) \\
0, & \text{otherwise}
\end{cases}
\end{equation}
where we define the effective attack duration as $\tau_{\text{eff}} = \tau_{\text{atk}} + \tau_{\text{recon}}$, representing the total time for one round of attack and reconnaissance. Similarly, the link attack can be described as
\begin{equation}
  \label{eq6}
E^\mathcal{L}_{ij}(t) = 
\begin{cases}
1, & 
\begin{aligned}
& (i, j) \in \mathcal{E}(t) \\
& \land~ f_i(t) = f_j(t) = f_{\text{atk}}(t) \\
& \land~ t \in [(n\!-\!1)\tau_{\text{eff}}, n\tau_{\text{eff}})
\end{aligned} \\
0, & \text{otherwise}
\end{cases}
\end{equation}

\subsection{Moving Target Defense Mechanisms}\label{section3.3}
\begin{figure*}[t]
  \centering
  \subfigure[Leader Switching]{
    \begin{minipage}[b]{0.3\textwidth}
      \includegraphics[width=\textwidth]{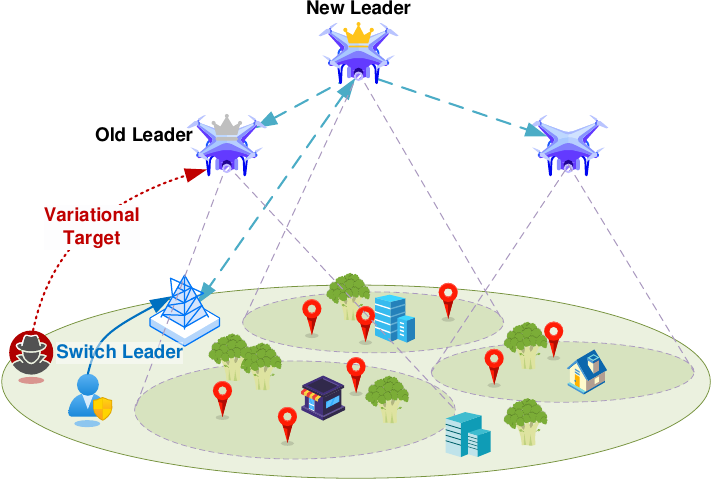}
    \end{minipage}
  }
  \hspace{1mm}
  \subfigure[Route Mutation]{
    \begin{minipage}[b]{0.3\textwidth}
      \includegraphics[width=\textwidth]{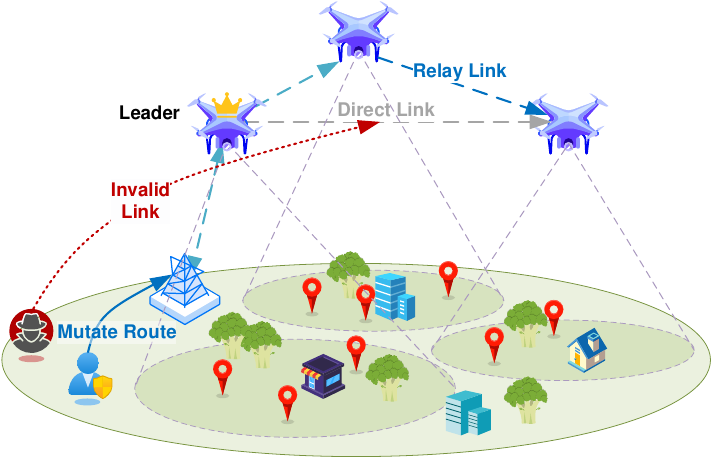}
    \end{minipage}
  }
  \hspace{1mm}
  \subfigure[Frequency Hopping]{
    \begin{minipage}[b]{0.3\textwidth}
      \includegraphics[width=\textwidth]{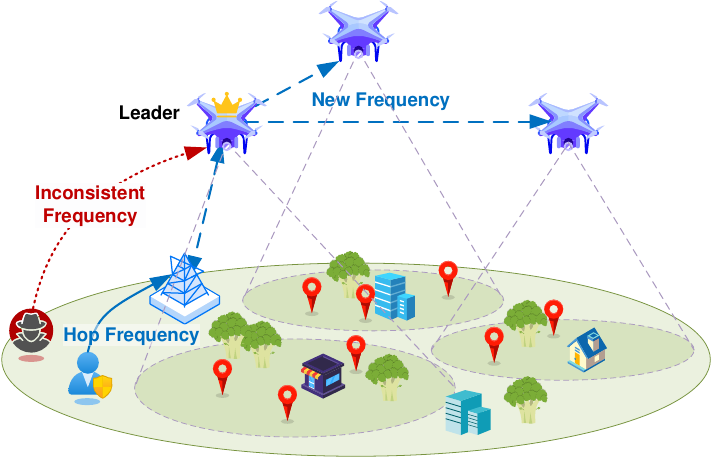}
    \end{minipage}
  }

\caption{\textcolor{black}{Overview of the proposed lightweight and coordinated Moving Target Defense (MTD) mechanisms for UAV swarm networks. (a) \textbf{Leader Switching}: Dynamically reassigns the logical leader role among UAVs, thereby enhancing resilience by preventing persistent targeting of a single node/link; (b) \textbf{Route Mutation}: Alters communication paths by selecting alternative relay UAVs, enabling critical messages to bypass compromised links and maintain connectivity under attack; (c) \textbf{Frequency Hopping}: Simultaneously changes the communication frequency channels of all UAVs and the ground control station, increasing uncertainty for attackers and disrupting their ability to sustain effective DoS attacks.}}
  \label{MTD_mechanisms}
\end{figure*}

Traditional defense mechanisms provide attackers with a stable attack surface, allowing them to gradually accumulate knowledge and optimize their strategies. To break this asymmetry and increase the attacker's uncertainty and cost, MTD has emerged as a proactive paradigm that dynamically shifts the system's attack surface, thereby disrupting the attacker's reconnaissance and exploitation process~\cite{celdran2024rl}.

While many MTD techniques, such as IP address randomization and virtual machine migration, have been successfully deployed in traditional network environments with elastic resources, their high computational and communication overhead makes them unsuitable for resource-constrained UAV swarms. Moreover, the real-time requirements and dynamic topology of UAV networks further limit the applicability of existing MTD solutions.

To address these challenges, we propose a suite of lightweight and coordinated MTD mechanisms specifically tailored for UAV swarm networks. These mechanisms are selected for their ability to be executed rapidly, with minimal resource consumption, and without disrupting the primary mission of the UAV swarm. The goal is to create a moving target for attackers, making it difficult for them to maintain effective attacks or predict the system's defensive responses. Fig.~\ref{MTD_mechanisms} provides an overview of the specific MTD-based mitigation mechanisms adopted in this work.

\subsubsection{Leader Switching}
As we discussed in Section~\ref{section3.2.2}, the leader UAV and the communication link between the GCS and the leader are often the primary targets for attackers, particularly those employing greedy strategies to maximize disruption. To counteract this vulnerability, we leverage the inherent dynamicity of the swarm to implement leader switching as a form of higher-level MTD mechanism.

Leader switching designates a new UAV as the swarm leader when the current leader is persistently targeted or isolated by attacks. It is important to note that leader switching is a logical (virtual) role reassignment that can be executed rapidly and with minimal overhead and does not require any physical movement within the swarm. This mechanism only involves updating the logical leader role in the existing control framework, requiring negligible computational resources and no additional communication overhead, thus having minimal impact on the physical formation or flight of the UAVs.

Let $\mathbb{L}^\mathcal{N}(t)$ denote the swarm leader at time $t$. The system can switch to a new leader chosen from the set of eligible UAVs after the leader switching occurs, such that
\begin{equation}
  \label{eq7}
  \mathbb{L}^\mathcal{N}(t+\tau^{\mathbb{L}}_{\text{exec}}) = i, \quad i \in \mathcal{V} \setminus \left\{0,\, \mathbb{L}^\mathcal{N}(t)\right\}
\end{equation}
Here, $i$ cannot be $0$ (the GCS) or the current leader and $\tau^{\mathbb{L}}_{\text{exec}}$ denotes the execution time that the leader switching mechanism takes. After the switch, the new leader takes over the responsibility of receiving control commands from the GCS and disseminating them to the follower UAVs through the flying Ad-Hoc network (FANET). 


\subsubsection{Route Mutation}
Route mutation is a mechanism that dynamically alters the communication paths within a network by changing the routing of data packets between nodes. In the context of UAV swarm networks, route mutation refers to the process of reconfiguring the communication topology so that messages can be delivered through different relay UAV nodes or links, rather than following a fixed, predictable path. 

Usually, the command is transferred from the GCS to the leader, and is then forwarded to the follower UAVs. However, when a direct communication link is blocked by an attack, we exploit route mutation that selects an alternative relay UAV to forward messages. For example, when the communication link between the current leader $\mathbb{L}^\mathcal{N}(t) = i$ and a follower UAV $j$ is under attack, then the relay UAV $r \in \mathcal{V} \setminus \{0, i, j\}$ should satisfy the following rules
\begin{equation}
  \label{eq8}
   E^\mathcal{L}_{ir}(t)=0 \land ( r, j ) \in \mathcal{E}(t)
\end{equation}

This operation dynamically reconfigures the communication graph in real time, allowing the command to be forwarded through the relay UAV and successfully delivered to the follower, thus mitigating the impact of the targeted link attack, i.e., $E^\mathcal{L}_{ij}(t+\tau^{\mathbb{R}}_{\text{exec}})=0$, where $\tau^{\mathbb{R}}_{\text{exec}}$ is the time for updating the routing table. 

By leveraging existing routing protocols and requiring only simple updates, this mechanism is lightweight, relying solely on local computations without introducing significant communication overhead. When there is no ongoing attack or when the attack fails to produce a substantial impact, the system can further optimize network performance by restoring the relay-based communication path back to a direct link. This reduces communication latency and improves overall efficiency without compromising the defense capability. 

\subsubsection{Frequency Hopping}
Despite the effectiveness of the aforementioned MTD mechanisms in mitigating DoS attacks, there remains a significant risk that a determined attacker could continuously target the leader UAV or the critical communication link between the GCS and the leader. Such persistent attacks may paralyze the entire UAV swarm's communication, rendering the swarm unable to coordinate or execute its patrol and mission tasks. Therefore, it is essential to introduce an additional defense mechanism that can further disrupt the attacker's ability to maintain effective defense and ensure the continuity of swarm operations.

Frequency hopping is a proactive MTD technique that addresses this challenge by dynamically and periodically changing the communication frequency channels used by all UAVs and the ground control station. At each defense opportunity, if the agent chooses the frequency hopping action, the swarm can switch from the last frequency to a new frequency selected from the available set $\mathcal{F}$, such that
\begin{equation}
  \label{eq9}
  f_i(t+\tau^{\mathbb{F}}_{\text{exec}}) \neq  f_i(t), \quad \forall i \in \mathcal{V},\quad f_i(t+\tau^{\mathbb{F}}_{\text{exec}}), f_i(t) \in \mathcal{F}
\end{equation}
where $\tau^{\mathbb{F}}_{\text{exec}}$ denotes the time cost of frequency hopping between all UAVs and the GCS. This approach significantly increases the uncertainty for attackers, as they must continuously detect and adapt to the new frequency to sustain their attack.

In practical UAV swarm deployments, frequency hopping can be implemented in a lightweight and coordinated manner. Each UAV and the ground control station maintain a synchronized hopping sequence, which can be generated using a pseudo-random algorithm or a shared secret key and be performed rapidly to minimize communication disruption against persistent and sophisticated DoS attacks.

\subsection{Problem Formulation}
For the UAV swarm network, its objective is to defend against the malicious DoS attacker. The UAVs need to choose the proper defensive actions to optimize their performance and cost in a distributed setting. Therefore, we formulate the optimization problem as
\begin{equation}
  \label{eq10}
  \begin{split}
  \textbf{P1:} \Big(&\min_{(i,j)\in \mathcal{E}(t)} \sum\limits_{t=1}^T E^\mathcal{N}_i(t)+E^\mathcal{L}_{ij}(t), \max_{t \in \mathbb{T}} |\mathcal{E}(t)|,\\
  & \min_{i \in \mathcal{V} \setminus \{0\}} \sum\limits_{t=1}^T |\mathbf{p}_i(t) - \mathbf{p}_i^*(t)|, \min_{i \in \mathcal{V} \setminus \{0\}} \sum\limits_{t=1}^T C^i(t) \Big)\\
  \textit{s.t. } &\textbf{C1:} \ Eqs.~(\ref{eq1}),(\ref{eq4}),(\ref{eq5}),(\ref{eq6}),(\ref{eq8})\\
  &\textbf{C2:}\ 0 \leq \theta_i(t) < 2\pi, \quad \forall i \in \mathcal{V} \setminus \{0\}, \forall t \in \mathbb{T}\\
  &\textbf{C3:}\ r\omega \leq v_i(t) \leq v_{\max}, \quad \forall i \in \mathcal{V} \setminus \{0\}, \forall t \in \mathbb{T}\\
  &\textbf{C4:}\ f_i(t) = f_j(t) \in \mathcal{F},\quad \forall i, j \in \mathcal{V}, \forall t \in \mathbb{T}\\
  &\textbf{C5:}\ 0 < n\tau_{\text{eff}} \leq T, \quad \forall n \in \mathbb{N}
  \end{split}
\end{equation}
where $|\mathcal{E}(t)|$ is the cardinality of this set and $C^i(t)$ is the defense cost of UAV $i$ at each time slot. Intuitively, the multi-objective optimization problem (\textbf{P1}) has the following objectives: (i) The first part aims to minimize the total number of effective node and link attacks over time, which is represented by the sum of $E^\mathcal{N}_i(t)$ and $E^\mathcal{L}_{ij}(t)$. (ii) The second part aims to maximize the network connectivity, which is represented by the maximum number of active communication links $|\mathcal{E}(t)|$ at any time step. (iii) The third part aims to minimize the deviation of each UAV from its ideal formation position, which is represented by the sum of the distance between each UAV's position and its ideal position $\mathbf{p}_i^*(t)$. (iv) Finally, the fourth part aims to minimize the cumulative defense cost, which is represented by the sum of defense costs $C^i(t)$ for all UAVs.

The constraints ensure that the UAV swarm operates within the defined parameters and have the following practical meanings: (i) Constraints \textbf{C1} enforce physical and communication limits, such as minimum separation between UAVs, attack effectiveness, and valid communication links. (ii) Constraint \textbf{C2} ensures that the heading angle of each UAV is within the range $[0, 2\pi)$, reflecting feasible orientation in 2D space. (iii) Constraint \textbf{C3} restricts the speed of each UAV to be within the patrol speed $r\omega$ and the maximum speed $v_{\max}$, ensuring safe and energy-efficient flight. (iv) Constraint \textbf{C4} ensures that all UAVs operate on the same frequency channel at any time step, which is essential for maintaining communication connectivity. (v) Finally, constraint \textbf{C5} ensures realistic attack cycles by stipulating that the total duration of any number of attacks remains within the mission time, where the effective duration of a single attack round $\tau_{\text{eff}}$ is the sum of the attack period and the reconnaissance period.

\subsection{Optimization Analysis}
It is worth noting that the multi-objective optimization problem above might not be solved by using traditional methods since obtaining accurate knowledge of the system dynamics and the attacker's behavior is impractical in real-world scenarios. Therefore, in this paper, we explore a DRL-based approach, leveraging the fitting capability of neural networks for approximating probability distributions and generating defense policies. 

To formalize the defense decision problem, we adopt a Partially Observable Markov Decision Process (POMDP) framework. A POMDP is particularly suitable for our UAV defense scenario because: (i) each UAV can only access local information rather than the complete system state, (ii) the environment changes in response to both defense and attack actions, and (iii) we need sequential decision-making under uncertainty. This formulation can be completely described through its state space, action space, transition probabilities, and reward function as follows.

\subsubsection{State Space}
Let $S_t$ denotes the global environment state at time step $t$, which is related to the state and action of the previous time step. In this scenario, it can be represented as
\begin{equation}
  \label{eq11}
  S_t = \left[ \mathbf{P}(t), \mathbf{V}(t), \mathbf{H}(t), \mathbf{L}(t), \mathbf{F}(t), \mathcal{E}(t) \right], \quad t\in \mathbb{T}
\end{equation}
where $\mathbf{P}(t), \mathbf{V}(t), \mathbf{H}(t)$, and $\mathbf{F}(t)$ denote the positions, velocities, headings, and frequencies of all UAVs at time $t$, respectively. For example, $\mathbf{P}(t) = [\mathbf{p}_1(t), \mathbf{p}_2(t), \ldots, \mathbf{p}_N(t)]$ and $\mathbf{F}(t) = [f_1(t), f_2(t), \ldots, f_N(t)]$. Moreover, $\mathbf{L}(t)$ is the binary indicator of the current leader UAV, which is the UAV that receives commands from the GCS and coordinates the swarm, such that $\mathbf{l}_i(t) = 1$ when $i = \mathbb{L}^\mathcal{N}(t)$, otherwise $0$. Finally, $\mathcal{E}(t)$ denotes the communication graph at time $t$, which captures the connectivity between UAVs based on their positions and frequencies (can be derived from Eq.~(\ref{eq1})).

The state of UAV $i$ at time step $t$ can be represented as $s_t^i$. However, each agent $i$ receives a local observation $o_t^i = O(S_t, i)$, which typically includes its own position, velocity, heading, the leadership indicator, its current frequency, and the local communication status. Formally, the observation space for agent $i$ can be defined as
\begin{equation}
  \label{eq12}
o_t^i = \left[ \mathbf{p}_i(t), \mathbf{v}_i(t), \mathbf{h}_i(t), \mathbf{L}(t), f_i(t), e_i(t)\right] 
\end{equation}
where $\mathbf{p}_i(t), \mathbf{v}_i(t), \mathbf{h}_i(t)$, and $f_i(t)$ are the position, velocity, heading, and the current frequency of agent $i$ at time $t$, respectively. For instance, when the UAV swarm patrols counterclockwise and each UAV always follows its ideal trajectory, the heading can be calculated as $h_i(t) = \theta_i(t) + \frac{\pi}{2}$. 

It should be noted that each agent is aware of the global leadership status $\mathbf{L}(t)$, as every UAV, regardless of whether it is the leader, needs to receive control commands from the current leader. In addition, the local communication status $e_i(t)$ can include information such as whether it can receive commands from the GCS (as a leader) or the current leader UAV (as a follower). This information is crucial for the agent to make informed decisions about its defense actions, which can be represented as
\begin{equation}
  \label{eq13}
  e_i(t) = \mathbb{I}\left[\, (0, i) \in \mathcal{E}(t) \;\; \text{or} \;\; (\mathbb{L}^\mathcal{N}(t), i) \in \mathcal{E}(t) \,\right]
\end{equation}

\subsubsection{Action Space}
At each decision point, each UAV agent needs to choose a defensive action. The set of all possible actions for an agent forms its action space. We assume that the agent makes defense decisions in an independent manner, implying that it may decide to take different actions on specific UAVs. At each time step, an agent $i$ can execute one of three MTD mechanisms as described in Section~\ref{section3.3}. Hence, we can define the action space for any state in $S_t$ as
\begin{equation}
  \label{eq14}
  A_t=[\tilde{A}_t^{\mathbb{L}}, \tilde{A}_t^{\mathbb{R}}, \tilde{A}_t^{\mathbb{F}}], \quad t\in \mathbb{T}
\end{equation}

\revise{Specifically, $\tilde{A}_t^{\mathbb{L}}=\{\mathbf{l}_1(t),\mathbf{l}_2(t),...,\mathbf{l}_N(t)\}$ denotes the action of leader switching at time step $t$. If a switch is performed, exactly one element $\mathbf{l}_i(t)$ will be 1, designating UAV $i$ as the new leader. If no switch occurs, all elements of $\tilde{A}_t^{\mathbb{L}}$ will be 0.} Similarly, $\tilde{A}_t^{\mathbb{R}}=\{a_{t,1}^{\mathbb{R}},a_{t,2}^{\mathbb{R}},...,a_{t,N}^{\mathbb{R}}\}$ denotes the actions of route mutation for all UAVs, where $a_{t,n}^{\mathbb{R}}=1$ represents that the $i$-th UAV plays the role of relay node whereas $a_{t,n}^{\mathbb{R}}=-1$ means canceling its role. Although the action space is formally defined for all agents, the initiation of global MTD mechanism (i.e., frequency hopping) is a privileged operation executed only by the current leader agent $\mathbb{L}^\mathcal{N}(t)$. When the leader's policy selects frequency hopping, the corresponding command is broadcast to all follower agents for synchronized execution. We define it as $\tilde{A}_t^{\mathbb{F}}=\{a_{t,1}^{\mathbb{F}},a_{t,2}^{\mathbb{F}},...,a_{t,N}^{\mathbb{F}}\}$, where $\tilde{A}_t^{\mathbb{F}}= [1,1,\ldots,1]$ triggers the change of the communication frequency for all UAVs and the GCS. This ensures that while all agents learn a comprehensive policy, only the leader's decision on global matters is enacted, thus resolving the coordination challenge.

Furthermore,  It is worth noting that a zero value indicates no action, that is, this MTD mechanism will not be executed at this time. Therefore, the action selected by agent $i$ at time step $t$ can be represented as
\begin{equation}
  \label{eq15}
  a_t^i = \left[\mathbf{l}_i(t), a_{t,i}^{\mathbb{R}}, a_{t,i}^{\mathbb{F}}\right] \in A_t, \quad i \in \mathcal{V} \setminus \{0\}
\end{equation}

\subsubsection{Transition Probability}
\revise{The state transition function $P(S_{t+1}|S_t, A_t)$ defines the probability of transitioning to state $S_{t+1}$ given the current state $S_t$ and the joint action $A_t$. In our simulated environment, the system dynamics are deterministic. This means that for any given state and action pair, the next state is uniquely determined. Formally, the transition probability can be expressed as}
\begin{equation}
  \label{eq16}
  P(s'|s,a) \in \{0,1\}, \quad \forall s,s' \in S_{t+1}, a \in A_t
\end{equation}

\revise{For instance, if the defender takes no action (i.e., all elements in $A_t$ are zero) on all UAVs and the attacker continues with the same malicious requests, the system remains in its current state, i.e., $P(s'|s,a)=1$ when $s'=s$ (and $0$ otherwise). Conversely, if the defender initiates a global frequency hopping when there is no ongoing attack, the frequency components of the next state are directly determined by this action, resulting in a predictable state transition with $P(s'|s,a)=1$ for a specific next state $s'$.}

\subsubsection{Reward Function}
In a multi-agent setting, each agent $i$ receives a reward based on its actions and the current state of the environment. In our scenario, we design to encourage the agents to maintain formation, ensure communication connectivity, and effectively respond to attacks while minimizing costs associated with defense actions. The reward function for agent $i$ at time step $t$ can be expressed as
\begin{equation}
  \label{eq17}
  r_t^i=\alpha R_{\text{com}}^i(t)+\beta R_{\text{form}}^i(t)-\zeta C^i(t)-\eta P_{\text{atk}}^i(t)-\xi P_{\text{vel}}^i(t)
\end{equation}
where $\alpha, \beta, \zeta, \eta$, and $\xi$ are weighting coefficients. These coefficients are crucial for balancing the trade-offs between mission objectives (e.g., connectivity, formation) and operational costs (e.g., defense actions, attack penalties). Their final values were determined through an empirical tuning process, as detailed in Section V-A. The components are defined as follows:

First, the reward of communication connectivity is defined by $R_{\text{com}}^i(t)=e_i(t)$, which is a binary indicator (defined in Eq.~(\ref{eq13})) that represents whether agent $i$ is connected to the leader or the GCS at time step $t$. Second, the reward of formation is defined by $R_{\text{form}}^i(t)=1-|\mathbf{p}_i(t)-\mathbf{p}_i^*(t)|/\delta$, which captures the deviation of agent $i$ from its ideal position $\mathbf{p}_i^*(t)$, normalized by the maximum allowed deviation threshold $\delta$ (defined in Section III-A).

Then, we define the cost of defense actions for agent $i$ as $C^i(t)=\sum_{j=1}^N \mathbf{l}_j(t)+a_{t,i}^{\mathbb{R}}+a_{t,i}^{\mathbb{F}}$. This cost function is composed of three parts: (i) The first term $\sum_{j=1}^N \mathbf{l}_j(t)$ represents the cost associated with the leader switching action. Since leader switching is a global action that affects the entire swarm's coordination, its cost is accounted for across all agents. (ii) The terms $a_{t,i}^{\mathbb{R}}$ and $a_{t,i}^{\mathbb{F}}$ are binary indicators (defined in Eq.~(\ref{eq15})) representing whether agent $i$ executes a route mutation or participates in a frequency hop, respectively. The penalty for being under effective attack is defined as $P_{\text{atk}}^i(t)=E^\mathcal{N}_i(t) + \frac{1}{2} E^\mathcal{L}_{i\mathbb{L}^\mathcal{N}(t)}(t)$, which captures the impact of node and link attacks on agent $i$. Finally, the penalty for excessive velocity is defined as $P_{\text{vel}}^i(t) = \frac {v_i(t) - v_{\text{pat}}}{v_{\max} - v_{\text{pat}}}$ which penalizes agent $i$ if its speed exceeds the allowed patrol speed.

The reward function effectively transforms the multi-objective optimization problem \textbf{P1} into a single-objective learning goal. By structuring the reward as a weighted sum of terms corresponding to the objectives in \textbf{P1} where maximization goals are positive rewards and minimization goals are penalties, we convert the problem into maximizing a single cumulative reward. The coefficients control the trade-offs between competing goals like maximizing security and minimizing cost.

\section{Federated Multi-Agent Deep Reinforcement Learning Framework}\label{section4}
To enable scalable, robust, and proactive defense in UAV swarm networks, we design a federated multi-agent deep reinforcement learning (FMADRL) framework. In this framework, each UAV is equipped with a local policy agent that learns to select MTD actions based on its own observations, while periodically participating in federated parameter aggregation to accelerate learning and improve generalization.

\subsection{Framework Overview}
The proposed FMADRL framework consists of $N$ distributed agents, each deployed on a UAV in the swarm, and a central aggregator (e.g., the GCS) responsible for federated parameter aggregation. 

During execution, each agent $i$ interacts with the environment based on its local observation $o_t^i$ (as defined in Eq.~(\ref{eq12})), selects an action $a_t^i$ from the action space $A_t$ (see Eqs.~(\ref{eq14}) and (\ref{eq15})), receives a reward $r_t^i$ (see Eq.~(\ref{eq17})), collects local experience, and updates its policy network parameters. The proposed framework allows the UAV swarm to learn adaptive and coordinated strategies against DoS attacks in a distributed and resource-efficient manner while ensuring scalability.

After a fixed number of episodes, agents upload their local model parameters to the aggregator, which computes a global average and redistributes the updated parameters back to the agents. This process enables collaborative learning without sharing raw data, thus reducing communication overhead.

\subsection{Local Policy Optimization}
We select a policy gradient (PG) based approach for local policy optimization due to its specific advantages in addressing the UAV defense problem. First, PG methods naturally learn a stochastic policy, which aligns perfectly with the MTD principle by introducing unpredictability into the UAV swarm. This makes it significantly harder for an attacker to counter the defensive actions. Second, PG methods are well-suited for the complex, multi-dimensional action space of our problem and generally offer more stable convergence properties compared to value-based methods like deep Q-network (DQN), which is crucial for learning a reliable policy in our dynamic environment.

\revise{In detail, deep neural network (DNN) models are utilized to build the learning agent in the local scenario.} Each agent maintains a policy network $\pi_{\theta_i}(a|o)$ parameterized by $\theta_i$. At each time step $t$, the agent receives a local observation $o_t^i$ and samples an action $a_t^i$ from the policy as
\begin{equation}
  \label{eq18}
  a_t^i \sim \pi_\theta(a_t^i | o_t^i)
\end{equation}

The policy updating procedure adjusts the parameter $\theta_i$ to improve the expected long-term cumulative reward of the agent $i$ for action $a_t^i$ given state $s_t^i$. To evaluate how good an action is in a given state, we use an action-value function (or Q-function), which estimates the expected total return per action. The Q-function is formulated as
\begin{equation}
  \label{eq19}
  Q^\pi(s, a) = \mathbb{E} \left[ \Omega_t^i | s=s_t^i, a=a_t^i \right]
\end{equation}
where $\Omega_t^i$ is the expected return for agent $i$ at time step $t$, which can be computed as the discounted sum of future rewards as
\begin{equation}
  \label{eq20}
  \Omega_t^i = \sum_{t=1}^{T} \gamma^{t-1} r_{t}^i
\end{equation}
where $\gamma \in [0, 1]$ is the discount factor that balances immediate and future rewards. Thus, we can also have the Q-function as
\begin{equation}
  \label{eq21}
  Q^\pi(s, a) = \mathbb{E} \Big[ r(s,a)+ \gamma \mathbb{E}_{a'\sim \pi} \left[Q^\pi(s', a') \right]\Big]
\end{equation}
where $r(s,a)$ is the immediate reward for taking action $a$ in state $s$, and $s'$ is the next state after taking action $a$.

The objective of policy learning is to develop an optimal policy $\pi_{\theta_i}^*$ that maps sequences to actions to maximize the objective function as 
\begin{equation}
  \label{eq22}
  \pi_{\theta_i}^* = \arg\max_\pi J(\theta_i)
\end{equation}
where $J(\theta_i)$ is the expected long-term cumulative reward of agent $i$, defined as
\begin{equation}
  \label{eq23}
  J(\theta_i) = \mathbb{E}_{s \sim d^{\pi_{\theta_i}},\, a \sim \pi_{\theta_i}} \left[ Q^{\pi_{\theta_i}}(s, a) \right]
\end{equation}
where $d^{\pi_{\theta_i}}$ is the state distribution under policy $\pi_{\theta_i}$.

Each agent stores its local experience tuples $(o_t^i, a_t^i, r_t^i, o_{t+1}^i)$ in a buffer $\mathcal{D}_i$. After collecting sufficient experience, the agent performs a policy gradient update, which adjusts the policy network's parameters $\theta_i$ in the direction that increases the expected reward. The gradient of the function should be calculated with respect to as follows
\begin{equation}
  \label{eq24}
  \nabla_{\theta_i} J(\theta_i) = \mathbb{E}_{\pi_{\theta_i}} \left[ \nabla_{\theta_i} \log \pi_{\theta_i}(a_t^i | o_t^i) Q^{\pi_{\theta_i}}(s_t^i, a_t^i) \right]
\end{equation}
where $\nabla_{\theta_i} \log \pi_{\theta_i}(a_t^i | o_t^i)$ is the score function that measures how much the policy changes with respect to the action taken. This gradient can be estimated using Monte Carlo sampling or temporal difference methods.  

For optimizing the objective function, the policy parameters are updated by minimizing the following loss function as
\begin{equation}
  \label{eq25}
  L(\theta_i) = -\mathbb{E}_{\pi_{\theta_i}} \left[ \log \pi_{\theta_i}(a_t^i | o_t^i) Q^{\pi_{\theta_i}}(s_t^i, a_t^i) \right]
\end{equation}
where the negative sign indicates that we want to maximize the expected return. To encourage exploration and prevent the agent from getting stuck in a suboptimal strategy, we add an entropy regularization term. The modified loss function becomes
\begin{equation}
  \label{eq26}
  L(\theta_i) = -\mathbb{E}_{\pi_{\theta_i}} \left[ \log \pi_{\theta_i}(a_t^i | o_t^i) \hat{R}_t^i - \mu H(\pi_{\theta_i}(\cdot | o_t^i)) \right]
\end{equation}  
where $\hat{R}_t^i$ is the normalized return, and $H(\cdot)$ is the entropy regularization term to encourage exploration. The entropy coefficient $\mu$ controls the trade-off between exploiting the current best-known actions and exploring new ones to discover potentially better strategies. A higher $\mu$ encourages more randomness in action selection, while a lower value leads to a more deterministic policy. Its value was determined empirically, as detailed in Section V-A.

In practice, the loss function can be expressed as
\begin{equation}
  \label{eq27}
  L(\theta_i) = -\frac{1}{T} \sum_{t=1}^{T} \log \pi_{\theta_i}(a_t^i | o_t^i) \cdot \hat{R}_t^i - \mu H(\pi_{\theta_i}(\cdot | o_t^i))
\end{equation}
The agent can update its policy parameters $\theta_i$ using gradient descent or adaptive optimization algorithms such as Adam.

\subsection{Federated Parameter Aggregation}
In our FMADRL framework, the aggregation of policy parameters across UAV agents is designed to maximize both learning efficiency and model personalization. 

\revise{After a number of local updates (e.g., $K_{\text{agg}}$ episodes), all agents upload their local parameters $\theta_i$ to the central aggregator.} Instead of naive averaging, we perform a reward-weighted aggregation of policy parameters. For each participating agent $i$, we compute a weight $w_i$ proportional to its recent average return over the most recent $M$ episodes, denoted as
\begin{equation}
  \label{eq28}
  \bar{R}_i^{(M)} = \frac{1}{M} \sum_{m=1}^{M} \left( \sum_{t=1}^{T} r_{t,m}^i \right)
\end{equation}
where $r_{t,m}^i$ is the reward received by agent $i$ at time step $t$ in the $m$-th most recent episode. The weight for agent $i$ is then calculated as
\begin{equation}
  \label{eq29}
  w_i = \frac{\bar{R}_i^{(M)}}{\sum\nolimits_{j=1}^N \bar{R}_j^{(M)}}
\end{equation}

To balance generalization and personalization, we only aggregate the parameters of the shared layers (e.g., initial hidden layers) while the final output layer is kept local. The boundary between them is determined by the principle of separating general knowledge from specialized decision-making, allowing each agent to personalize its final action-selection mapping based on its specific role and recent experiences. Thus, the global parameter for each shared layer is computed as
\begin{equation}
  \label{eq30}
  \theta^{\text{global}} = \sum\limits_{i=1}^N w_i \theta_i
\end{equation}

After receiving the updated shared parameters $\theta^{\text{global}}$, each agent $i$ synchronizes the shared layers of its local policy network. To further personalize the policy, each agent then performs $T_{\text{local}}$ steps of local fine-tuning using its own recent experience buffer $\mathcal{D}_i$. Specifically, the agent updates its full parameter set $\theta_i$ by minimizing the local loss function (e.g., Eq.~(\ref{eq27})). This process enables each agent to adapt the global model to its local environment and recent experiences, thus achieving a balance between collaborative learning and individual specialization.

\subsection{Policy Gradient-Based FMADRL Algorithm}

To provide a clear overview of the FMADRL framework for MTD deployment in defeating DoS mitigation, we summarize the overall training procedure in the proposed policy gradient-based FMADRL (PG-FMADRL) algorithm (see Algorithm~\ref{alg1}). This algorithm integrates the local policy optimization and federated parameter aggregation steps described in previous sections, enabling distributed UAV agents to collaboratively learn robust MTD strategies for DoS attack mitigation.

\subsubsection{Algorithm Description}
\revise{Algorithm~\ref{alg1} details the training process for our PG-FMADRL method. The procedure iterates for $K_{\max}$ episodes, alternating between local policy updates and periodic federated aggregation. In the local phase, each agent $i$ interacts with the environment, collecting experience tuples $(o_t^i, a_t^i, r_t^i, o_{t+1}^i)$ into its buffer $\mathcal{D}_i$. At the end of each episode, it updates its local policy parameters $\theta_i$ by minimizing the loss function $L(\theta_i)$ (Eq.~(\ref{eq27})) via policy gradient.}

\revise{The federated aggregation phase commences every $K_{\text{agg}}$ episodes, during which each agent contributes its local parameters $\theta_i$ and recent average reward $\bar{R}_i^{(M)}$ to the central aggregator. The aggregator then computes an improved global model $\theta^{\text{global}}$ using reward-weighted averaging (Eq.~(\ref{eq30})). Finally, agents synchronize their shared layers with $\theta^{\text{global}}$ and perform $T_{\text{local}}$ local fine-tuning steps to adapt the generalized policy to their specific context.}

\begin{algorithm}[tbp]
\caption{PG-FMADRL Algorithm}
\label{alg1}
Initialize global shared parameters $\theta^{\text{global}}$ \;
Initialize local policy parameters $\theta_i \leftarrow \theta^{\text{global}}$ \;
Initialize local experience buffer $\mathcal{D}_i \leftarrow \emptyset$ \;
\For{$k = 1$ to $K_{\max}$}{
    \For{$i = 1$ to $N$ \textbf{(in parallel)}}{
        Reset environment and observe initial $o_1^i$\;
        \For{$t=1$ to $T$}{
            Select action $a_t^i$ to execute it (Eq.~(\ref{eq18}))\;
            Receive $r_t^i$ (Eq.~(\ref{eq17})) and observe $o_{t+1}^i$\;
            Store $(o_t^i, a_t^i, r_t^i, o_{t+1}^i)$ in $\mathcal{D}_i$\;
        }
        Compute $\Omega_t^i$ (Eq.~(\ref{eq20})) and normalize it as $\hat{R}_t^i$\;
        Update $\theta_i$ by minimizing $L(\theta_i)$ (Eq.~(\ref{eq27})) using policy gradient (Eq.~(\ref{eq24}))\;
    }
    \If{mod ($k$, $K_{\text{agg}}$) = 0}{
        \For{$i = 1$ to $N$ \textbf{(in parallel)}}{
            Compute average return $\bar{R}_i^{(M)}$ over recent $M$ episodes (Eq.~(\ref{eq28}))\;
            Upload $\theta_i$ and $\bar{R}_i^{(M)}$ to the aggregator\;
        }
        Compute weights $w_i$ (Eq.~(\ref{eq29})) and aggregate parameters using Eq.~(\ref{eq30})\;
        \For{$i = 1$ to $N$ \textbf{(in parallel)}}{
            Synchronize shared layers: $\theta_i \leftarrow \theta^{\text{global}}$\;
            \For{$t=1$ to $T_{\text{local}}$}{
                Sample mini-batch from $\mathcal{D}_i$ and update $\theta_i$ (Eqs.~(\ref{eq24}) and (\ref{eq27}))\;
            }
        }
    }
}
\end{algorithm}

\subsubsection{Complexity Analysis}
During execution, each UAV agent makes decisions at every time step by performing a forward pass through its local policy network $\pi_{\theta_i}$, which is typically implemented as a DNN. For a policy network with $L_{\text{P}}$ layers and $N_n$ neurons in the $n$-th layer, the computational complexity of a single forward pass is $O(\sum_{n=1}^{L_{\text{P}}-1} N_n N_{n+1})$. The local policy update, based on the policy gradient method, involves both forward and backward passes, resulting in a per-update complexity of the same order.

In each episode, the agent interacts with the environment for $T$ time steps, leading to a per-episode complexity of $O(T \sum_{n=1}^{L_{\text{P}}-1} N_n N_{n+1})$. After every $K_{\text{agg}}$ episodes, the federated aggregation step computes the weighted average of $P$ shared parameters across $N$ agents, which has a complexity of $O(NP)$. Then, the local fine-tuning phase consists of $T_{\text{local}}$ gradient updates per agent, each with complexity $O(\sum_{n=1}^{L_{\text{P}}-1} N_n N_{n+1})$. Thus, the overall computational complexity is $O(NK_{\text{agg}}T \sum_{n=1}^{L_{\text{P}}-1} N_n N_{n+1} + NP + N T_{\text{local}} \sum_{n=1}^{L_{\text{P}}-1} N_n N_{n+1})$.

\section{Performance Evaluation}\label{section5}
This section evaluates the effectiveness and efficiency of our proposed PG-FMADRL method through extensive simulations. We first describe the simulation setup and baseline methods used for comparison. Then, we analyze the convergence properties of our algorithm, followed by a comprehensive assessment of its performance in terms of attack mitigation rate, recovery time, energy consumption, and defense cost under various attack scenarios. Finally, we assess the scalability of our approach compared to other SOTA methods.

\subsection{Simulation Setup}
We investigate a coverage area of $1 \times 1$ km serviced by a single GCS and multiple UAVs. The GCS is located at the center of the area, specifically at coordinates [500, 500, 0] m. Each UAV in our method is responsible for patrolling and monitoring the ground area, and all of them build a formation with a radius of 300 m and a height of 100 m. For our primary experiments (Sections~\ref{section5.2} to~\ref{section5.4}), we use a configuration with $N=5$ UAVs for a detailed and fair comparison across multiple metrics and baselines. To specifically address scalability, we then conduct an analysis in Section V-E with varying swarm sizes. The minimum separation distance between UAVs is set to 20 m, and the deviation threshold for maintaining formation is set to 40 m. The attack duration is set to 15 s, and the reconnaissance period is set to 5 s. Each MTD mechanism has an execution time of 1 s.

\revise{The reward coefficients $(\alpha, \beta, \zeta, \eta, \xi)$ and entropy coefficient $\mu$ were determined through an iterative tuning process. Our methodology prioritized critical objectives, setting a high weight for the attack penalty $\eta=2$ to focus on threat mitigation, followed by the velocity penalty $\xi=0.5$ to limit energy use. The remaining coefficients were fine-tuned around $[0, 1]$. Each configuration was evaluated based on learning stability, final converged reward, and attack mitigation rate.} The final values for all simulation parameters are summarized in Table~\ref{parameter} with reference to~\cite{lei2024multi,chai2024system,xu2025low}.

All the experiments were conducted on a workstation with 2.20 GHz Intel Xeon Gold 5220R, NVIDIA RTX A5000, 128 GB RAM, and Ubuntu 20.04 operating system. During the training phase, we trained the models for 2000 episodes and repeated the training process with random seeds ranging from 1 to 10 to ensure statistical reliability. In the testing phase, each method was evaluated over 10 episodes using random seeds from 1 to 100 to reduce the uncertainty of the environment.

\begin{table}[t]
  \caption{Key Parameters Used in the Simulation.}
  \centering
  \label{parameter}
  \begin{tabular}{@{}lc@{}}
    \toprule
    Parameter & Value or Range \\ \midrule
    Area size & $1000 \times 1000$ m$^2$ \\
    GCS position $\mathbf{p}_0$ & [500, 500, 0] m \\
    Patrol radius $r$ and height $h$ & 300 m, 100 m \\
    Number of UAVs $N$ & 5 \\
    Number of frequency channels $|\mathcal{F}|$ & 5 \\
    Communication range $R_c$ & 500 m \\
    Patrol speed $v_{\text{pat}}$ and 
    maximum speed $v_{\max}$  & 15 m/s, 20 m/s \\
    Minimum separation $d_{\min}$ & 20 m \\
    Deviation threshold $\delta$ & 40 m \\
    Attack duration $\tau_{\text{atk}}$ & 15 s \\
    Reconnaissance period $\tau_{\text{recon}}$ & 5 s \\
    MTD Execution time $\tau_{\text{exec}}^{\mathbb{L}}$, $\tau_{\text{exec}}^{\mathbb{R}}$, and $\tau_{\text{exec}}^{\mathbb{F}}$ & 1 s \\
    Time steps $T$ per episode & 50 \\
    Training episodes $K_{\max}$ & $2 \times 10^{3}$ \\
    Discount factor $\gamma$ & 0.99 \\
    Hidden layers & [64, 64] \\
    Learning rate (Adam)& $1 \times 10^{-3}$ \\
    Batch size & 128 \\
    Activation function & ReLU \\
    Memory size of $\mathcal{D}_i$ & $2 \times 10^{4}$ \\
    Federated aggregation interval $K_{\text{agg}}$ & 20 \\
    Reward averaging window $M$ & 20 \\
    Local fine-tune steps $T_{\text{local}}$ & 100 \\
    Reward coefficients $(\alpha, \beta, \xi, \eta, \zeta)$ & (0.5, 0.5, 1, 2, 0.5) \\
    Entropy coefficient $\mu$ & 0.01 \\
    \bottomrule
  \end{tabular}
\end{table}

To verify the effectiveness of the proposed PG-FMADRL algorithm, the following schemes are used as benchmarks and the performance is compared with that of the proposed algorithm, which are described as follows.

\begin{itemize}
  \item \textbf{WF-MTD}~\cite{tan2022wf}: It introduces rationality parameters to describe the learning capabilities of both the attacker and the defender. Then, it develops a Wright-Fisher process-based method for selecting the optimal MTD strategy.
  \item \textbf{RE-MTD}~\cite{zhou2025resource}: This method formulates the interaction between attacks and MTD deployment as a Markov decision process (MDP), and adopt a DQN algorithm to achieve a trade-off between effectiveness and overhead.
  \item \textbf{ID-HAM}~\cite{zhang2023how}: This scheme models an MDP to describe the MTD mutation process, and designs an advantage actor-critic algorithm to learn from scanning behaviors and slow down network reconnaissance intelligently.
  \item \textbf{DESOLATER}~\cite{yoon2021desolater}: It is a multi-agent deep reinforcement learning (MADRL)-based MTD technique that enables the agents to learn robustly in the presence of partial observations when defending against attacks.
\end{itemize}

It is worth noting that, although some of the baseline methods were originally proposed for general network environments or cloud networks rather than UAV-specific scenarios, we have carefully adapted and implemented all baseline methods within the UAV swarm context. Specifically, each baseline is configured to operate under the same UAV swarm system model as our proposed approach, and all utilize the same set of lightweight MTD mechanisms as described in Section~\ref{section3.3}. The original algorithms and decision-making frameworks of each baseline are retained, and all methods are trained and evaluated using the same simulation environment and attack scenarios to ensure consistency and fairness in the comparative evaluation.

\subsection{Convergence Analysis}\label{section5.2}
\begin{figure}[t]
  \centering
  \subfigure[Node attack]{
    \includegraphics[width=0.46\linewidth]{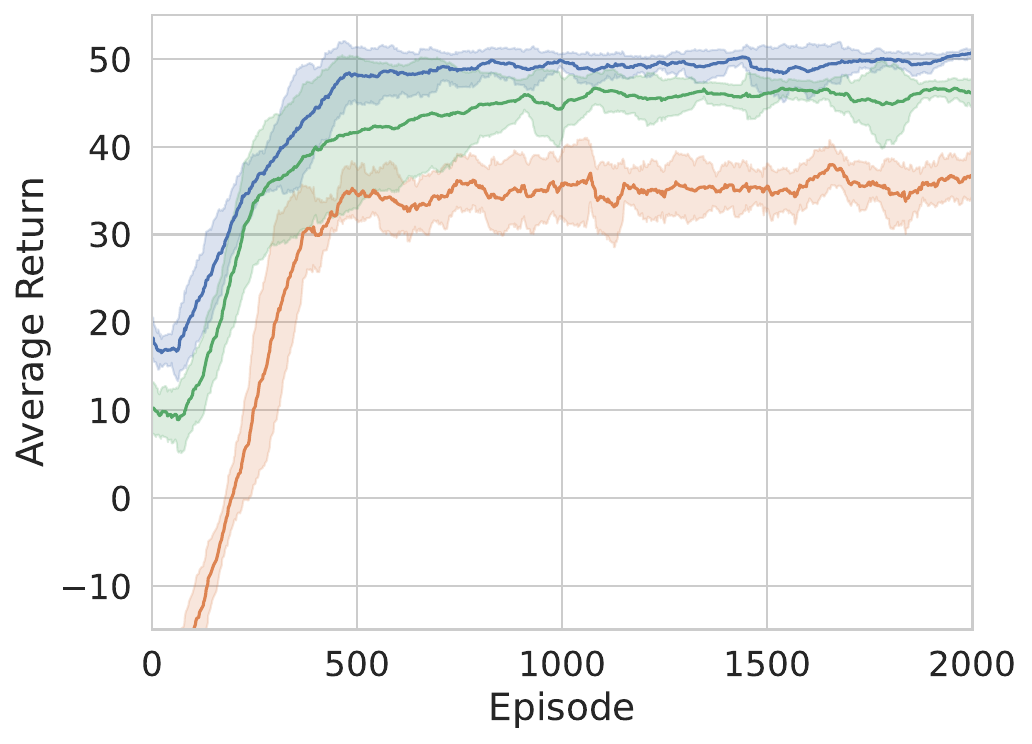}
  }
  \subfigure[Link attack]{
    \includegraphics[width=0.46\linewidth]{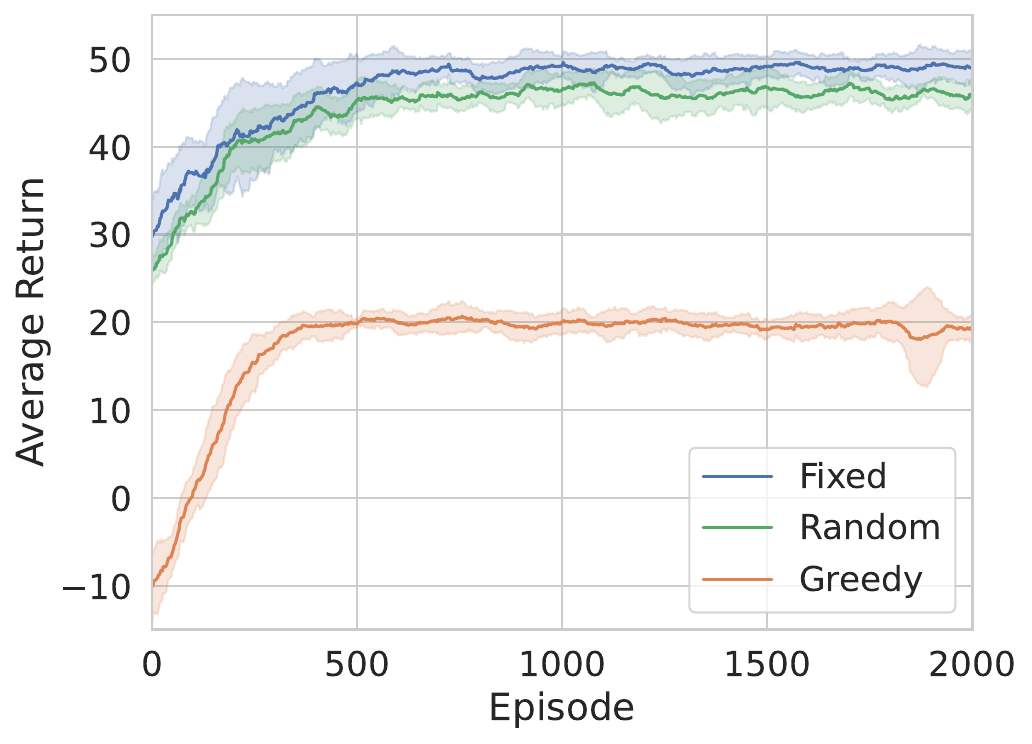}
  }
  \caption{The average return of the proposed PG-FMADRL method under different DoS attack types and strategies.}
  \label{Average_return}
\end{figure}

To evaluate the convergence of our proposed method, we measured the average reward under fixed, random, and greedy attackers in both link and node scenarios. In Fig.~\ref{Average_return}, the $x$-axis indicates the number of training episodes, while the $y$-axis presents the average reward at each episode. In all cases, the average return increases steadily with the number of episodes and reaches convergence at approximately episode 500, which indicates that the proposed method achieves consistently good defense performance across different attack scenarios.

As shown in Fig.~\ref{Average_return}, fixed and random attackers lead to relatively high average rewards. In both link-level and node-level scenarios, the fixed attacker achieves a return close to 50, while the random attacker yields approximately 47. The reason is that fixed and random attackers select their targets once, and the applied defense mechanisms remain effective throughout the episode. In contrast, greedy attackers continuously adjust their targets in response to the defense, which results in the decrease of reward.

Besides, the figure also indicates that the proposed method performs well under both link and node attack scenarios. For fixed and random attackers, the average return remains similar across the two scenarios, suggesting that these relatively simple strategies of attackers have limited impact regardless of the attack scenarios. However, a notable performance drop is observed under greedy attackers, where the return decreases from 35 in the node scenario to 20 in the link scenario. This result suggests that link attacks are more difficult to defend, likely due to their broader influence on system communication and resource availability.

In the end, the rapid and stable convergence of PG-FMADRL can be attributed to its federated learning framework, which enables UAV agents to share policy parameters and aggregate knowledge from diverse local experiences. This collaborative approach accelerates learning, helps avoid local optima, and ensures that the swarm quickly adapts to various attack strategies, especially in dynamic and partially observable environments.

\begin{table*}[t]
\caption{Attack Mitigation Rate under Different Attack Types and Strategies. The highest average mitigation rates (\textcolor{green!}{green}) and the lowest standard deviations (\textcolor{orange!}{orange}) for each attack are highlighted.}
\centering
\label{Mitigation_rate}
\begin{tabular}{@{}l|cc|cc|cc||cc|cc|cc@{}}
\toprule
\multirow{4}{*}{Defense Method} 
& \multicolumn{6}{c||}{Node Attack} 
& \multicolumn{6}{c}{Link Attack} \\ 
\cmidrule(lr){2-7} \cmidrule(lr){8-13}
& \multicolumn{2}{c|}{Fixed Attacker} 
& \multicolumn{2}{c|}{Random Attacker} 
& \multicolumn{2}{c||}{Greedy Attacker} 
& \multicolumn{2}{c|}{Fixed Attacker} 
& \multicolumn{2}{c|}{Random Attacker} 
& \multicolumn{2}{c}{Greedy Attacker} \\
\cmidrule(lr){2-13}
 & Avg. & StdDev. & Avg. & StdDev. & Avg. & StdDev. & Avg. & StdDev. & Avg. & StdDev. & Avg. & StdDev. \\ 
\midrule
WF-MTD     & 0.8974 & 0.0236 & 0.8706 & 0.0129 & 0.7192 & 0.0104 & 0.9684 & 0.0025 & 0.9465 & 0.0036 & 0.7721 & 0.0105 \\
RE-MTD     & 0.9413 & 0.0053 & 0.9276 & 0.0054 & 0.6535 & 0.0236 & 0.9823 & \cellcolor{orange!20}0.0015 & 0.9615 & \cellcolor{orange!20}0.0028 & 0.7360 & 0.0201 \\
ID-HAM     & 0.9897 & 0.0024 & 0.9964 & 0.0010 & 0.8646 & 0.0106 & 0.9405 & 0.0070 & 0.9598 & 0.0033 & 0.7625 & 0.0074 \\
DESOLATER  & 0.9966 & \cellcolor{orange!20}0.0004 & 0.9992 & 0.0004 & \cellcolor{green!20}1.0000 & \cellcolor{orange!20}0.0000 & \cellcolor{green!20}0.9969 & \cellcolor{orange!20}0.0015 & 0.9778 & 0.0057 & 0.7910 & 0.0081 \\
PG-FMADRL  & \cellcolor{green!20}0.9975 & \cellcolor{orange!20}0.0004 & \cellcolor{green!20}0.9999 & \cellcolor{orange!20}0.0000 & 0.9996 & 0.0002 & \cellcolor{green!20}0.9969 & \cellcolor{orange!20}0.0015 & \cellcolor{green!20}0.9782 & 0.0039 & \cellcolor{green!20}0.9367 & \cellcolor{orange!20}0.0014 \\
\bottomrule
\end{tabular}
\end{table*}

\subsection{Effectiveness and Efficiency Analysis}
\revise{To analyze the defense effectiveness and operational efficiency of our method, we first evaluate the attack mitigation rate under various scenarios. We then assess the system's resilience by measuring the recovery time required to restore connectivity after an attack. We compare our PG-FMADRL method against four representative MTD-based approaches, including WF-MTD, RE-MTD, ID-HAM, and DESOLATER.}

\begin{figure}[t]
  \centering
  \subfigure[Node attack]{
    \includegraphics[width=0.46\linewidth]{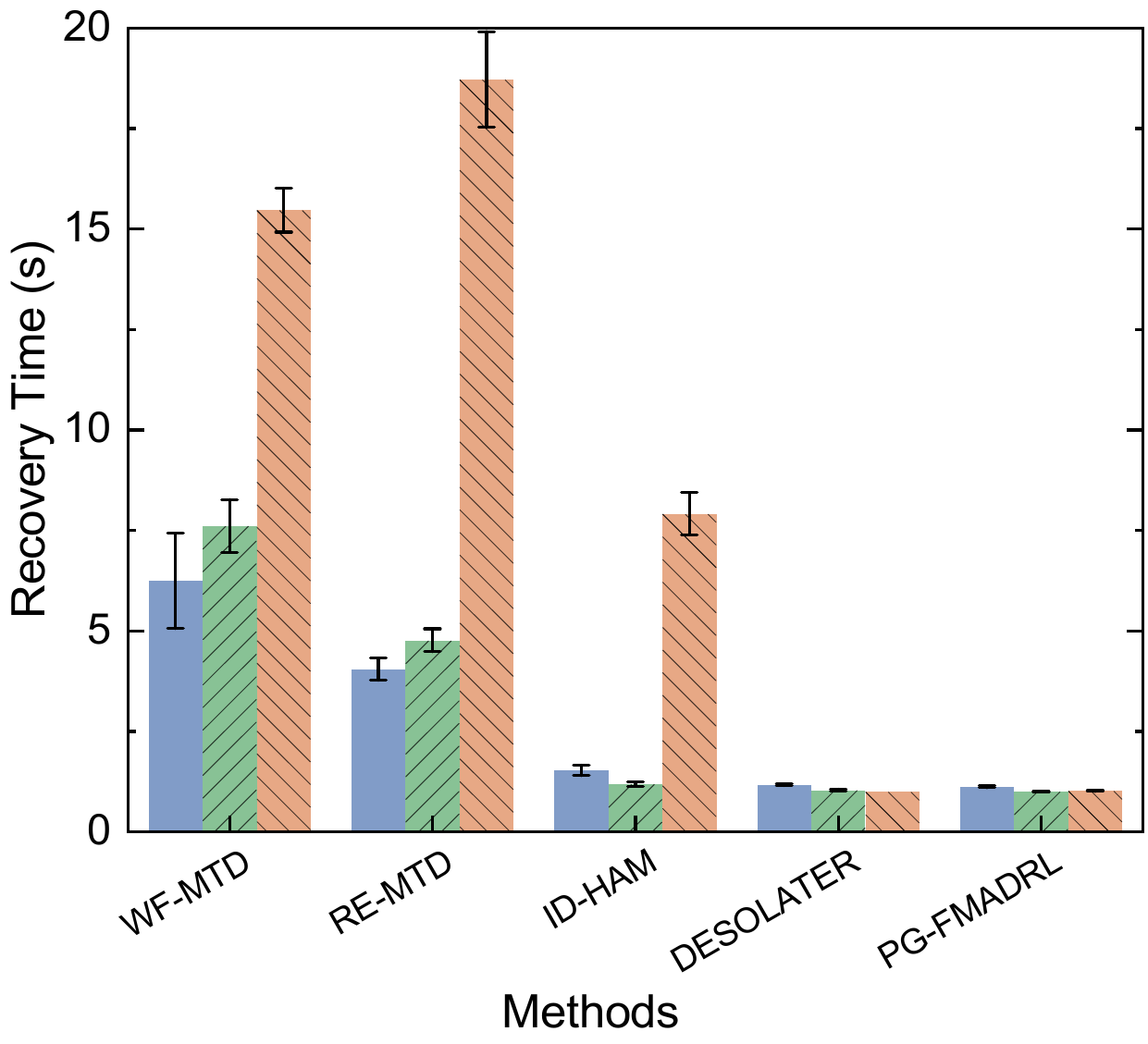}
  }
  \subfigure[Link attack]{
    \includegraphics[width=0.46\linewidth]{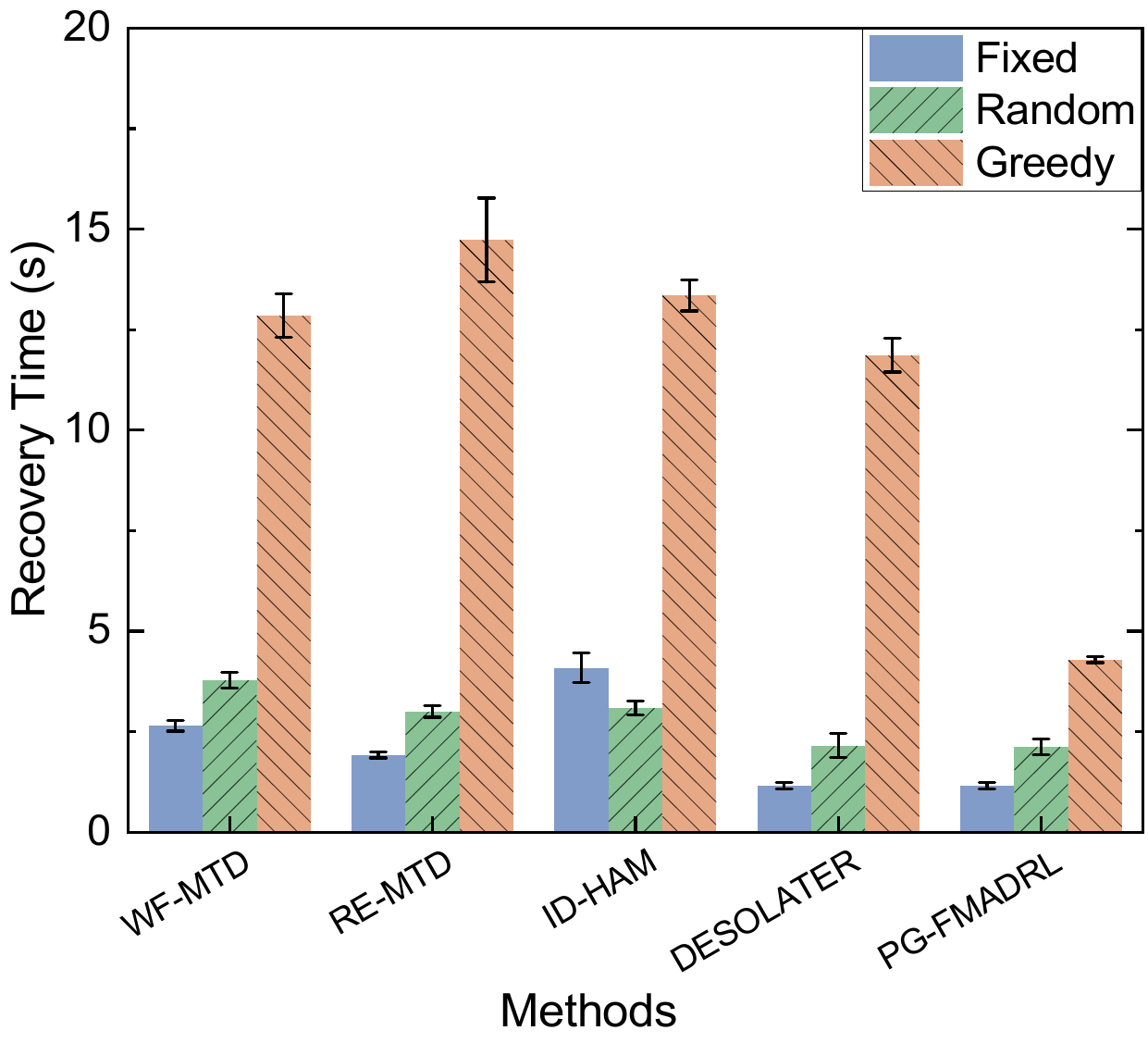}
  }
  \caption{The recovery time of the UAV swarm under different solutions for defeating DoS attacks with different types and strategies.}
  \label{Recovery_time}
\end{figure}

The defense attack mitigation rate is defined as the proportion of time steps within an episode during which attacks are successfully mitigated. Specifically, At each step, a heartbeat detection is performed for each UAV to compute its communication score, which is then used to assess connectivity. A UAV is considered disconnected if the score falls below a predefined threshold, indicating a defense failure at that step. The final attack mitigation rate is computed as the average proportion of UAVs that remain connected throughout the episode. The results are summarized in the Table~\ref{Mitigation_rate}.

We can see that PG-FMADRL consistently achieves the highest or near-highest attack mitigation rates across all attack scenarios. For node attack scenario, PG-FMADRL achieves highest attack mitigation rates of 0.9975 and 0.9999 against fixed and random attackers, respectively, while maintaining the lowest standard deviation among all compared methods. This indicates superior stability and robustness. In the link attack scenario, although the attack mitigation rates show a slight decline, the proposed method still achieves the highest values, reaching 0.9969 for fixed and 0.9782 for random attackers. These results demonstrate the method's strong and consistent defense effectiveness across different scenarios.

Notably, PG-FMADRL's advantage becomes most apparent against greedy attackers. While ID-HAM and DESOLATER rely on independent learning where each agent must discover effective strategies in isolation, our federated architecture acts as a knowledge accelerator. For instance, in the challenging greedy link attack scenario, PG-FMADRL reaches an attack mitigation rate of 0.9367, significantly outperforming the other methods, which achieve only 0.7910, 0.7625, 0.7360, and 0.7721, respectively. This is because our reward-weighted aggregation mechanism allows the entire swarm to rapidly learn from the experiences of the most successful agents those achieving high rewards, by effectively countering the attack. Our collaborative learning ensures that effective defense policies are quickly propagated and adopted across the swarm, leading to a more robust and coordinated response.

\revise{Beyond mitigation rate, we also assess the efficiency of each method using recovery time, defined as the duration required for the system to restore normal connectivity following a disruption. Fig.~\ref{Recovery_time} presents the recovery times of different defense methods.}

As we can see, in the node attack scenario, WF-MTD shows the longest recovery times under fixed and random attackers, reaching 6.2 and 7.6 seconds, respectively. Under greedy attacks, RE-MTD performs the worst, with a recovery time of 18.7 seconds. In contrast, our PG-FMADRL method dramatically reduces the recovery time by approximately 94.6\%, requiring about 1 second to restore a secure state. In the link attack scenario, the worst recovery times are observed for different methods depending on the attacker type. Such prolonged recovery durations leave the UAV swarm vulnerable for extended periods, increasing both operational risk and energy consumption. Although DESOLATER performs relatively well under greedy attacks in the node scenario, its recovery time increases significantly to 11.8 seconds in the link scenario. On the other hand, PG-FMADRL achieves the shortest recovery times across all attack scenarios and attacker types, enabling rapid mitigation and minimizing disruption.

In summary, the superior attack mitigation rates and reduced recovery times of PG-FMADRL stem from its reward-weighted aggregation and distributed policy learning. This enables UAVs to collaboratively reinforce successful defense behaviors in real time. Compared to other multi-agent methods such as DESOLATER and ID-HAM, this approach ensures that high-performing agents have greater influence on the global policy, leading to more robust and consistent defense against dynamic and adaptive attacks.

\begin{figure}[t]
  \centering
  \subfigure[Node attack]{
    \includegraphics[width=0.46\linewidth]{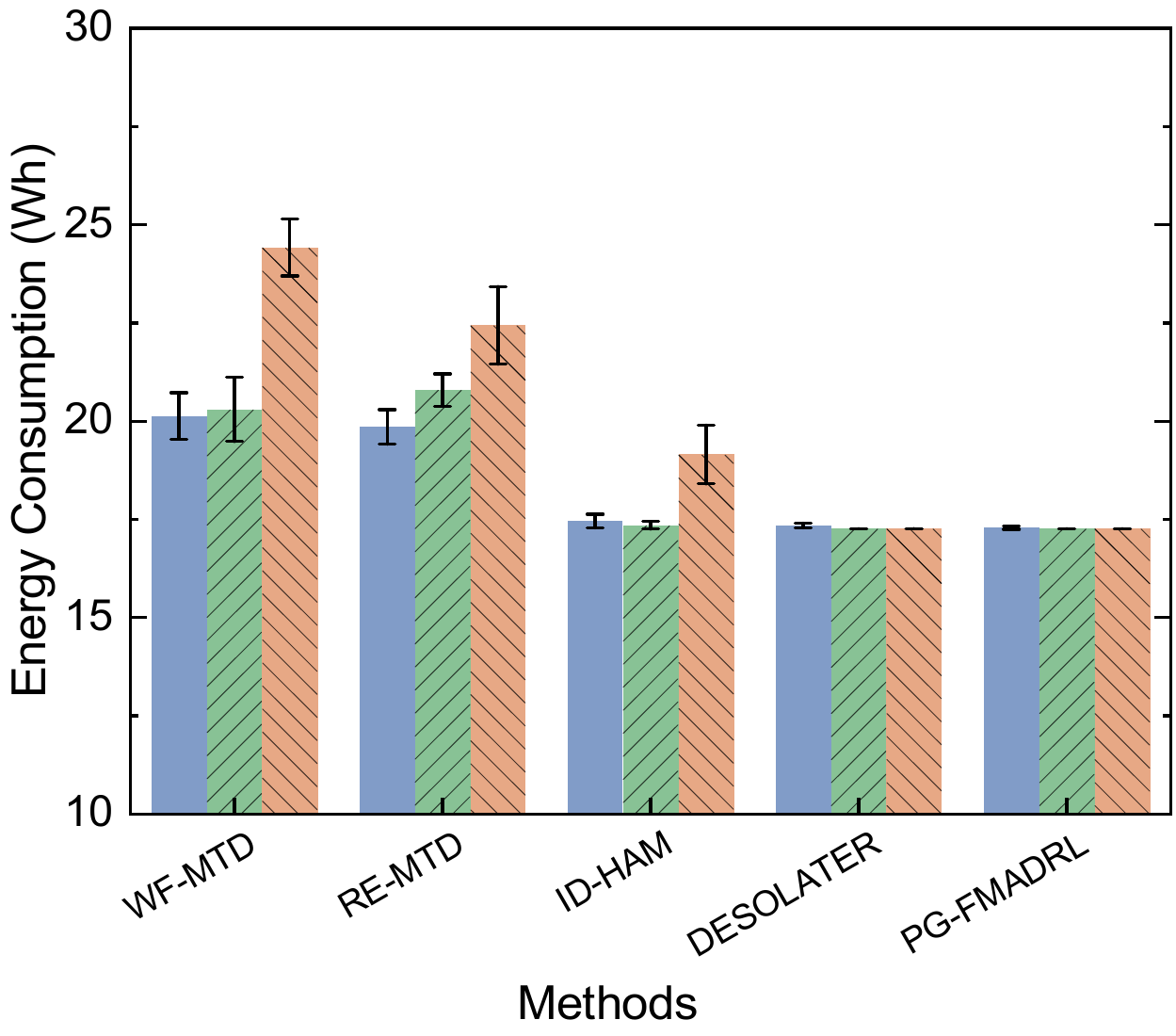}
  }
  \subfigure[Link attack]{
    \includegraphics[width=0.46\linewidth]{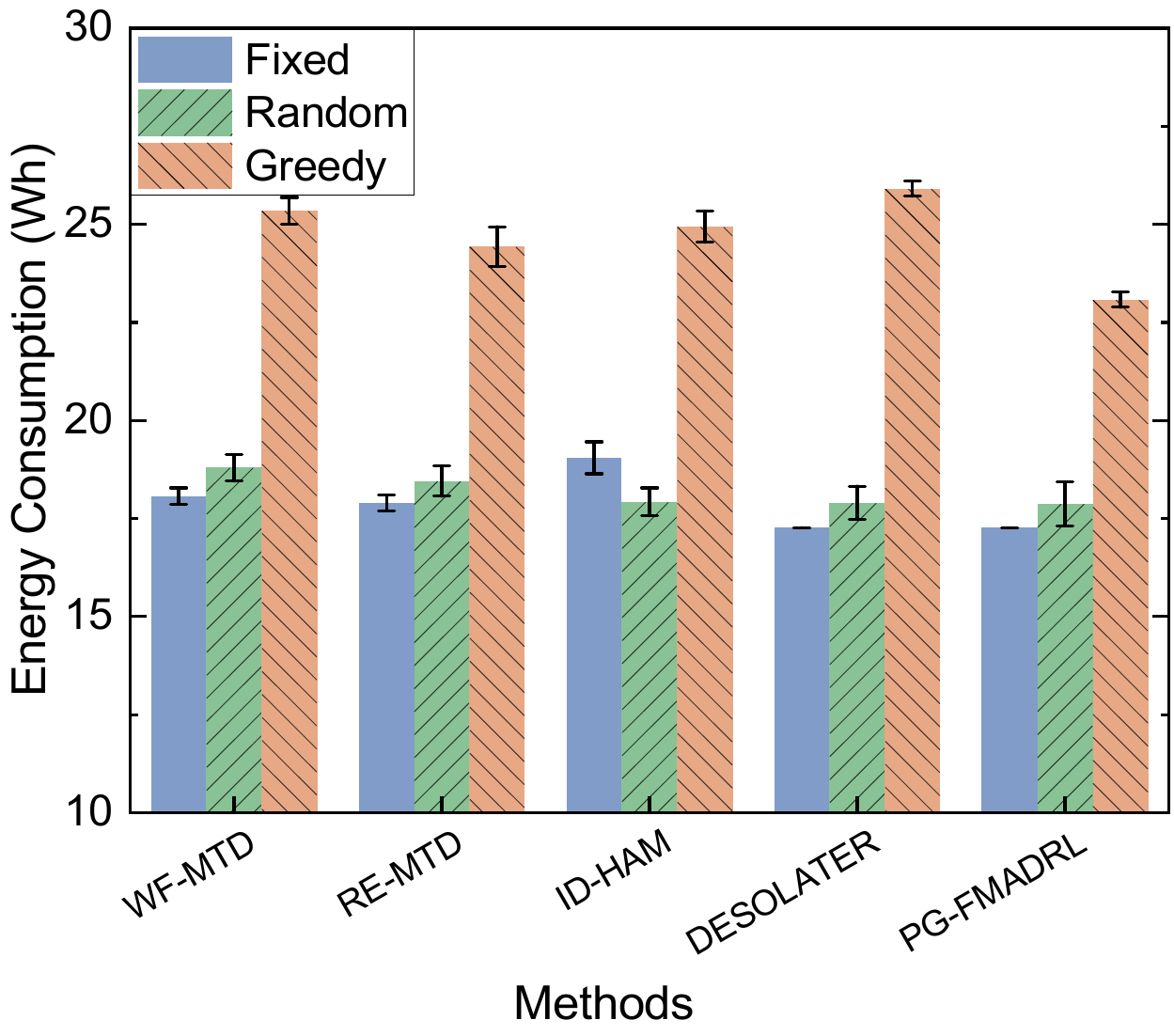}
  }
  \caption{The energy consumption of the UAV swarm under different solutions for defeating DoS attacks with different types and strategies.}
  \label{Energy_consumption}
\end{figure}

\subsection{Energy Consumption and Defense Cost Analysis}\label{section5.4}

\begin{figure}[t]
  \centering
  \subfigure[Node attack from fixed attacker]{
    \includegraphics[width=0.46\linewidth]{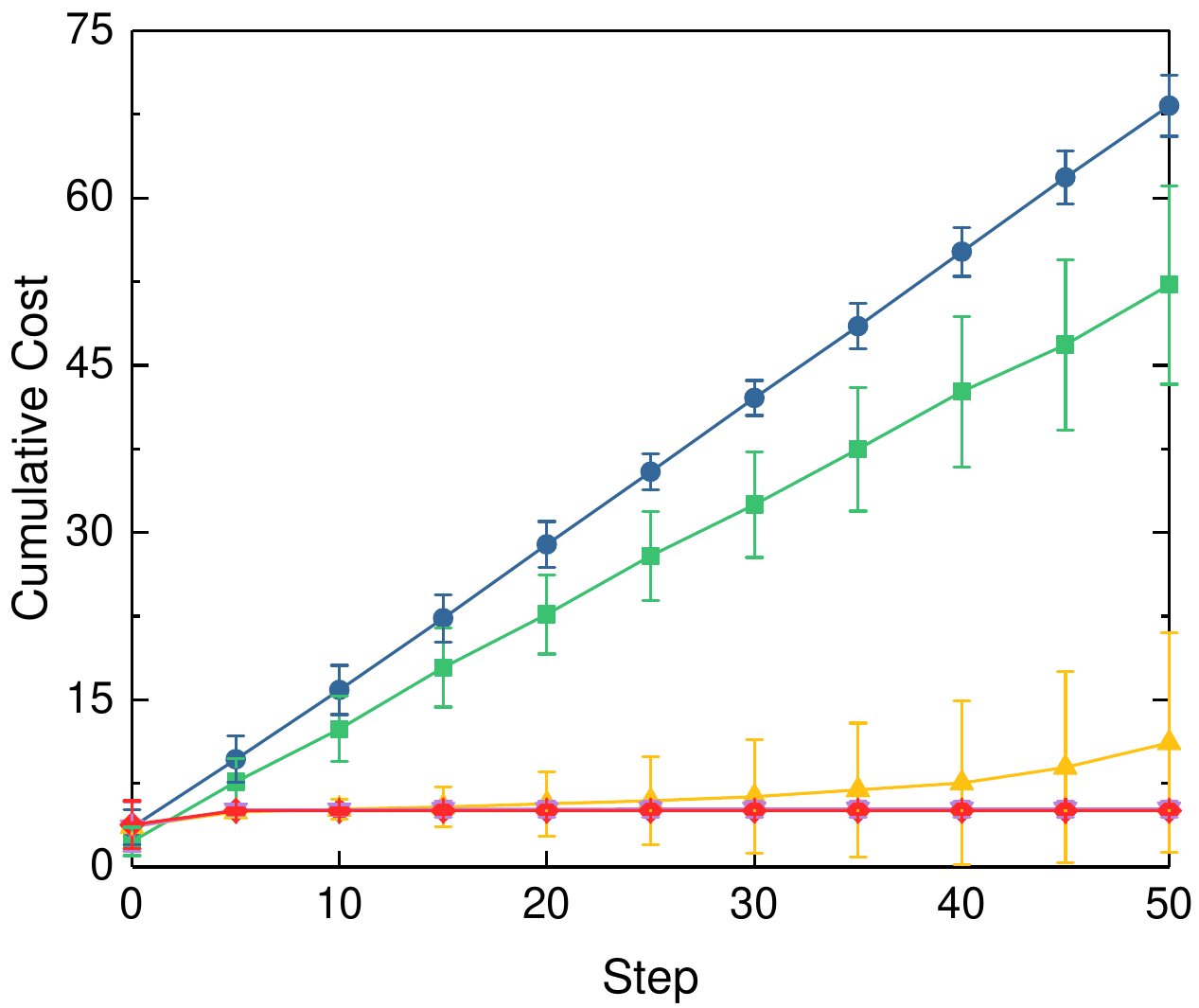}
  }
  \subfigure[Link attack from fixed attacker]{
    \includegraphics[width=0.46\linewidth]{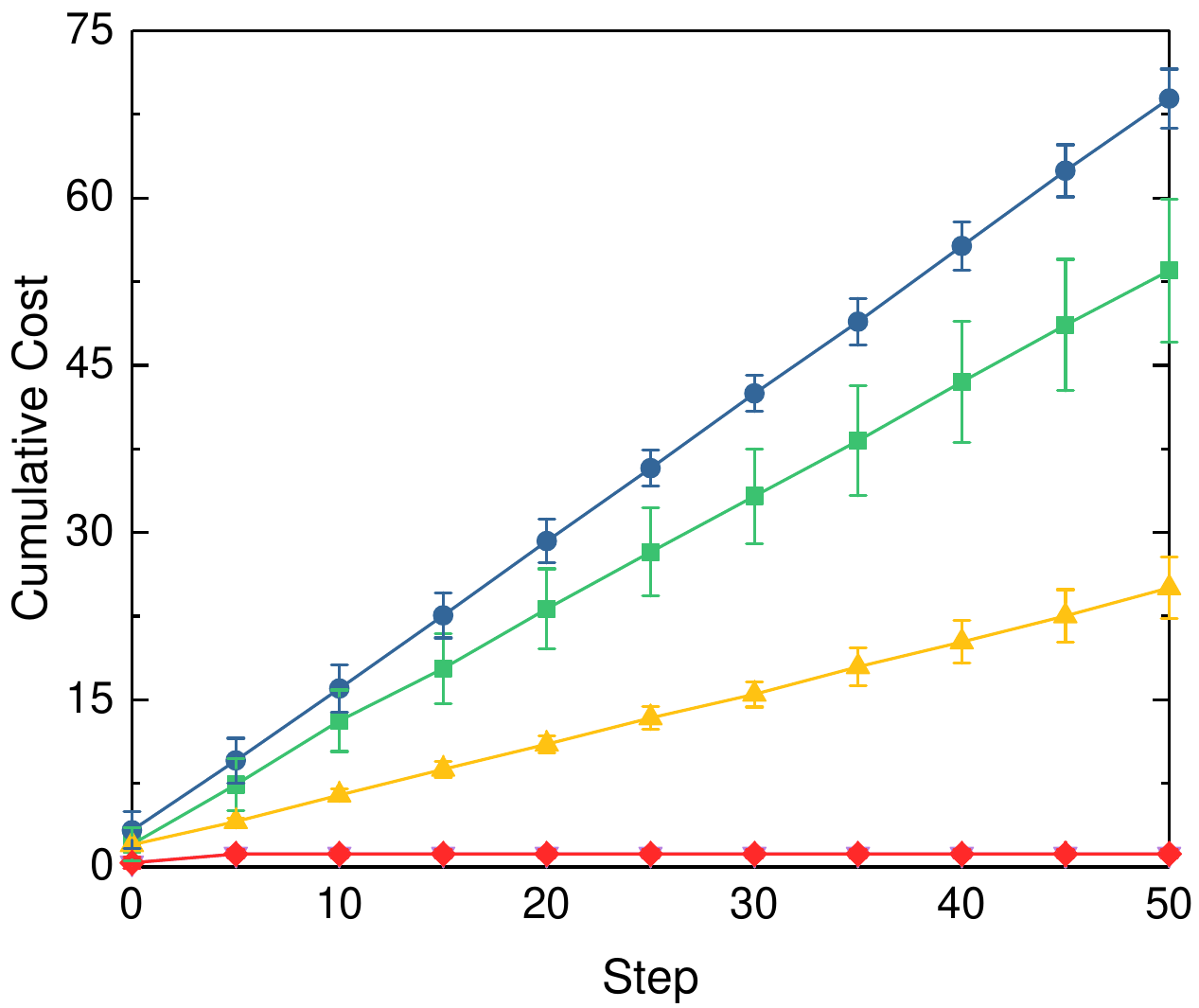}
  }
  \vspace{1mm}
  \subfigure[Node attack from random attacker]{
    \includegraphics[width=0.46\linewidth]{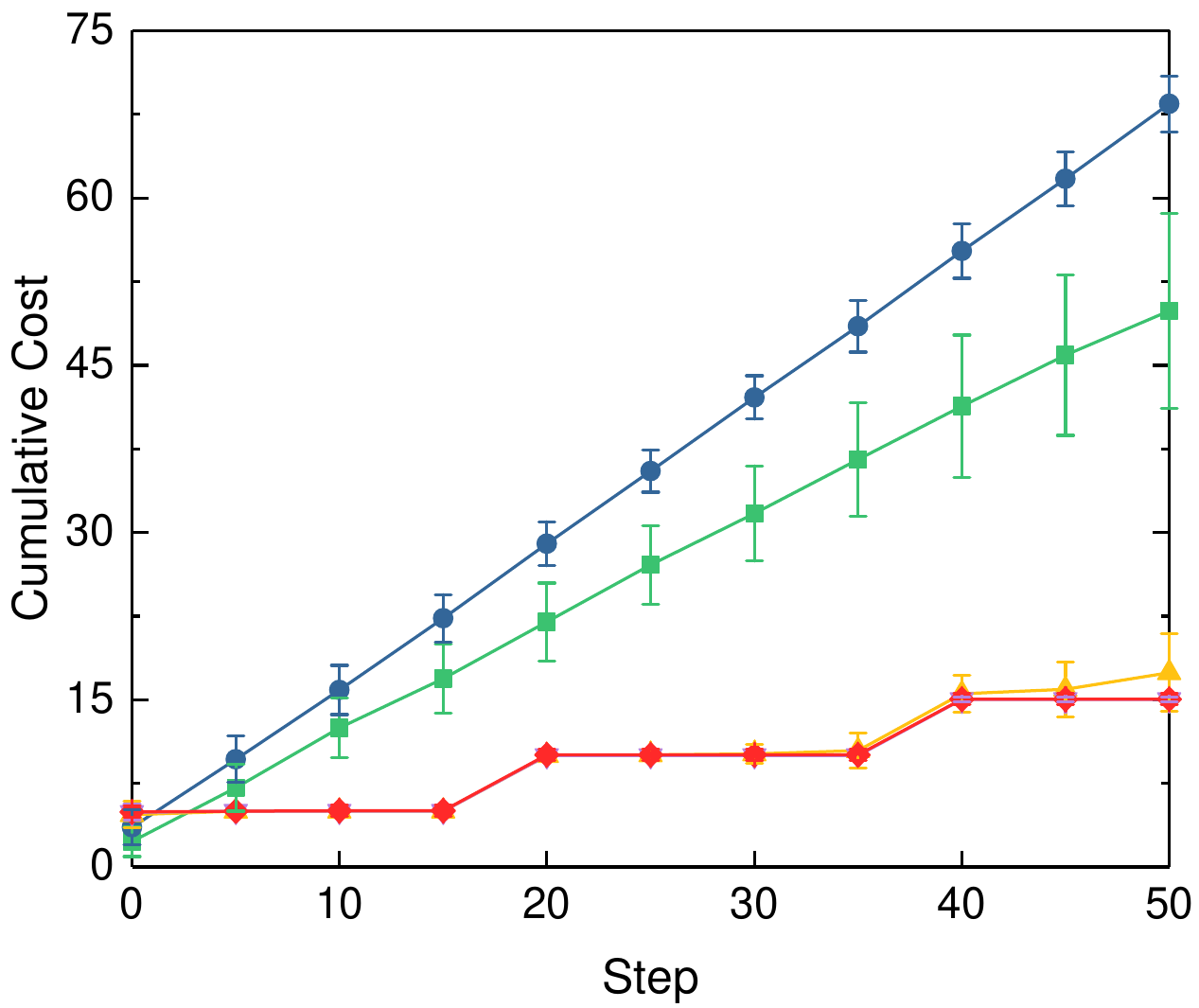}
  }
  \subfigure[Link attack from random attacker]{
    \includegraphics[width=0.46\linewidth]{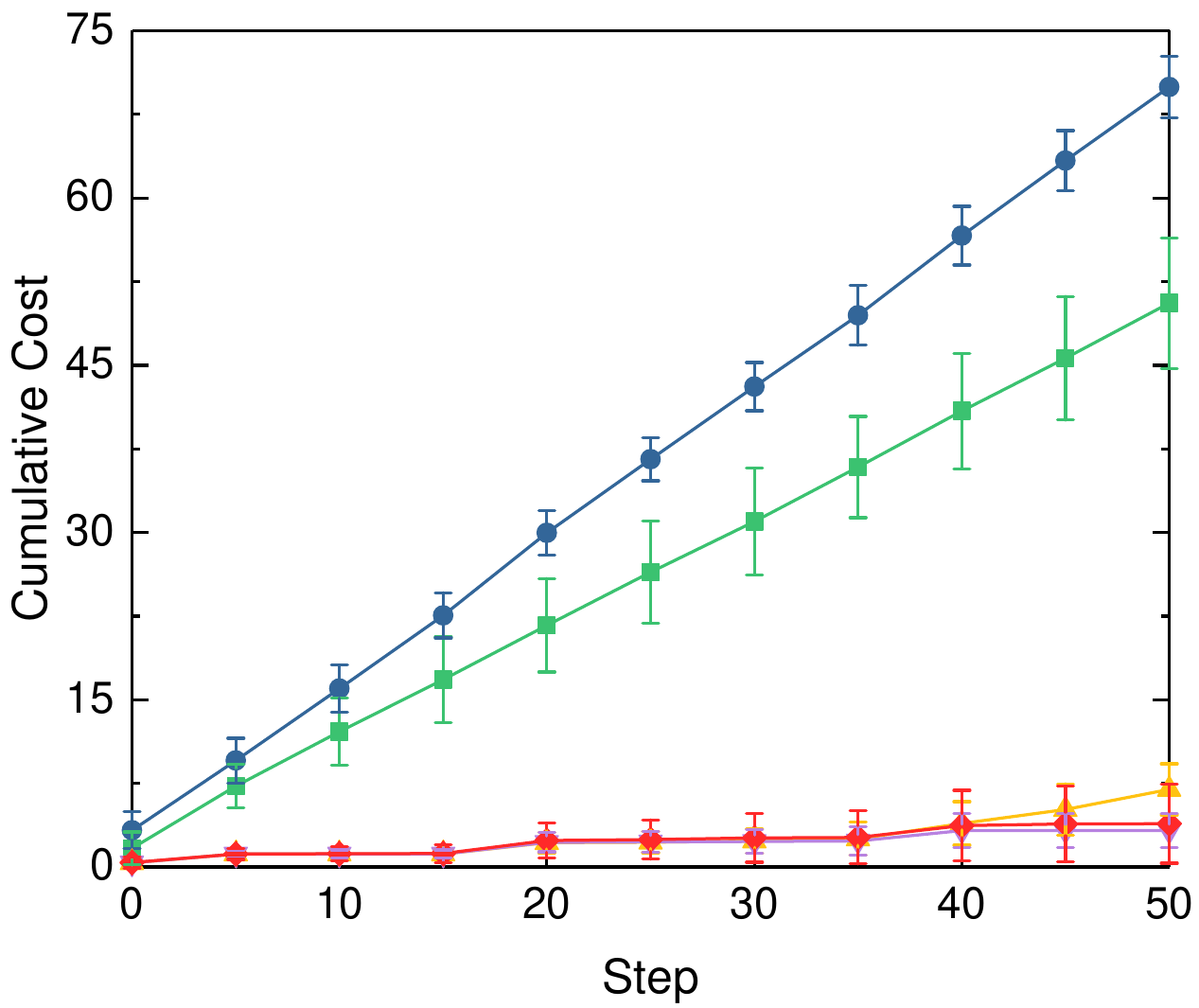}
  }
  \vspace{1mm}
  \subfigure[Node attack from greedy attacker]{
    \includegraphics[width=0.46\linewidth]{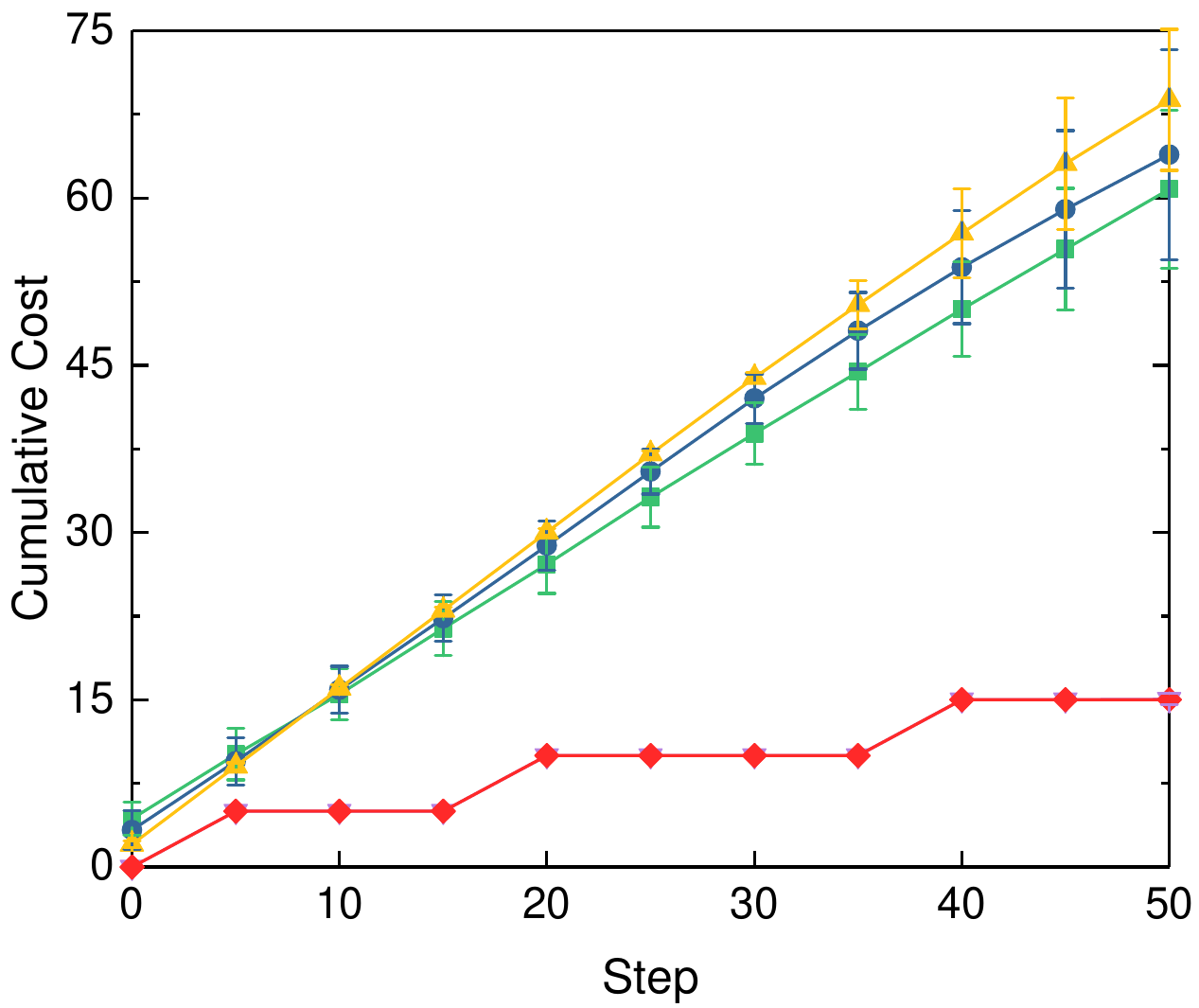}
  }
  \subfigure[Link attack from greedy attacker]{
    \includegraphics[width=0.46\linewidth]{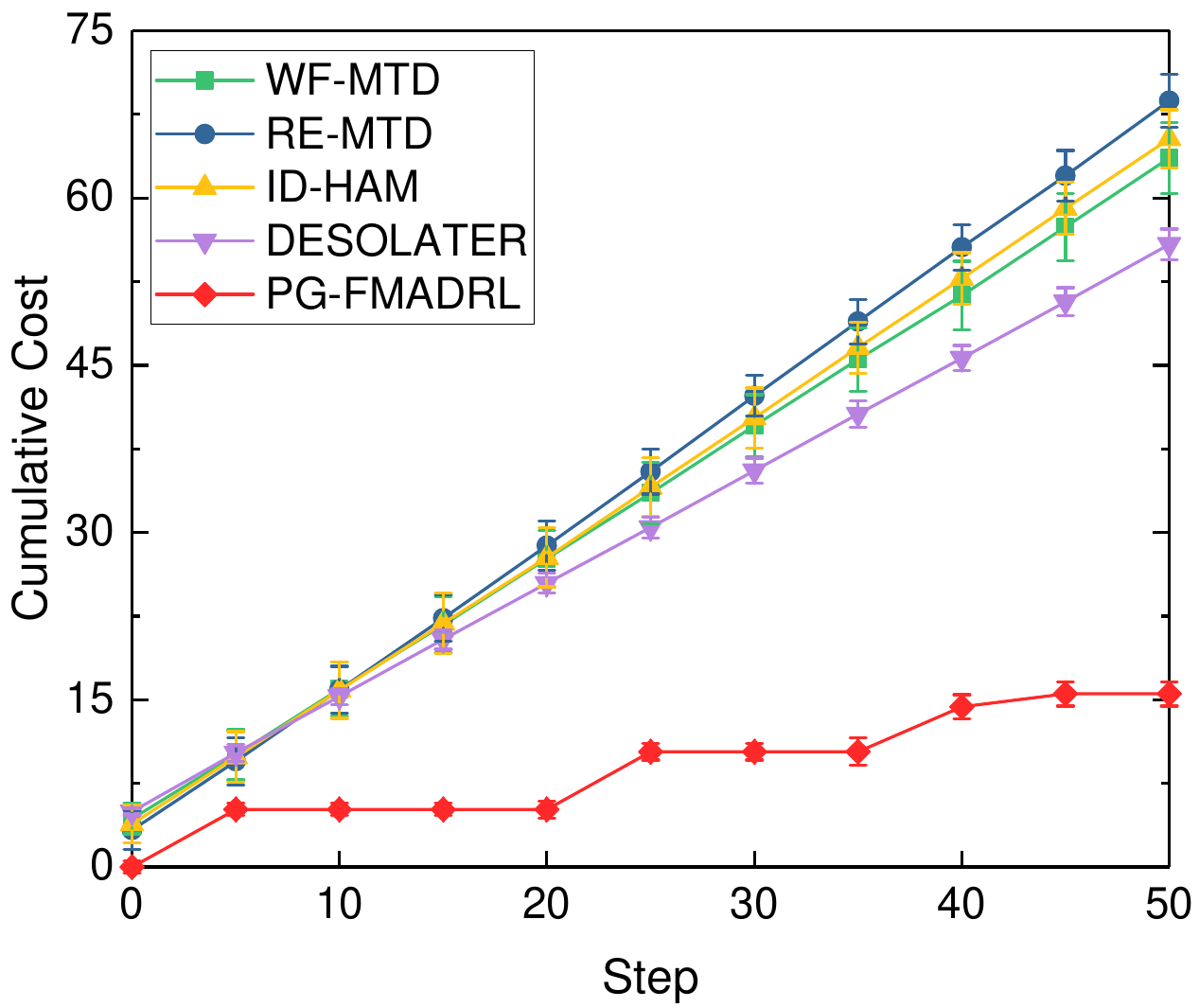}
  }
  \caption{The cumulative defense cost of the UAV swarm under different solutions for defeating DoS attacks with different types and strategies.}
  \label{Cumulative_cost}
\end{figure}

\revise{While the previous section established the effectiveness and efficiency of our method, a comprehensive evaluation must also consider the associated overhead. This section analyzes the operational costs from two perspectives, including the energy consumed by the UAVs during defensive maneuvers and the cumulative cost of executing MTD actions.}

First, we adopt a simplified model introduced by Liu \emph{et al.}~\cite{liu2017power} to express the power of a UAV according to its velocity, which includes three components, namely induced power, profile power, and parasite power as
\begin{equation}
P(v_i) = (c_1 + c_2) \cdot (mg)^{3/2} + c_3 \cdot v_i^3
\end{equation}
where $c_1$ and $c_2$ represent empirical coefficients related to induced and profile power ($c_1 = 2.8037\, (\mathrm{m}/\mathrm{kg})^{1/2}$, $c_2 = 0.3177\, (\mathrm{m}/\mathrm{kg})^{1/2}$), $c_3$ is the parasitic power coefficient ($c_3 = 0.0296\, \mathrm{kg}/\mathrm{m}$), $m$ is the mass of the UAV ($m = 1.283\, \mathrm{kg}$), and $g$ is the gravitational acceleration ($g = 9.8\, \mathrm{m}/\mathrm{s}^2$) with reference to~\cite{liu2017power}.

This model allows us to quantitatively compute the consumed power and compare the energy efficiency of each defense strategy, and the results are presented in Fig.~\ref{Energy_consumption}. From an overall perspective, the greedy attack strategy leads to significantly higher energy consumption for the UAV swarm compared to the fixed and random strategies. This is primarily because the greedy attacker exhibits greater attack intensity and adaptability, dynamically adjusting its targets in response to defense actions. As a result, the swarm is often forced to deviate from its optimal formation, causing UAVs to travel longer distances and thus to consume more energy. Nevertheless, PG-FMADRL consistently achieves the lowest energy consumption across all scenarios. Notably, under node attack conditions, it reduces energy consumption by 29.3\% compared to WF-MTD when facing greedy attackers, thereby effectively extending the patrol duration of the swarm. 

We can also observe that the energy consumption of DESOLATER is very close to our method under node attacks, which can be attributed to the fact that DESOLATER also employs a policy gradient-based approach. However, our methods enables more effective agent coordination by leveraging the federated learning framework. Consequently, under greedy attack strategies, PG-FMADRL achieves a 10.9\% reduction in energy consumption compared to DESOLATER, highlighting its effectiveness in collaborative defense and energy efficiency.

\revise{In addition to energy consumption, we analyze the cumulative defense cost, which reflects the resource expenditure from triggering MTD mechanisms.} As we can see in Fig.~\ref{Cumulative_cost}, the cumulative defense cost varies significantly across different methods and attack scenarios. It is evident that the proposed method consistently achieves the lowest cumulative cost, regardless of the attack type or attacker strategy. This demonstrates the superior efficiency of our approach in minimizing unnecessary defense actions while maintaining robust protection. Furthermore, baseline methods such as WF-MTD and RE-MTD incur significantly increasing costs, as they tend to trigger more frequent or redundant actions due to their lack of adaptive coordination. In contrast, for both node and link attacks from fixed and random attackers, PG-FMADRL maintains a nearly flat cost curve, indicating that the swarm can effectively mitigate attacks with minimal resource expenditure. Specifically, PG-FMADRL achieves a cumulative cost of 1.16 under fixed link attacks when the episode ends, having a significant reduction of 98.3\% compared to RE-MTD. 

\begin{figure*}[t]
  \centering
  \subfigure[$N$ = 10]{
    \begin{minipage}[b]{0.18\textwidth}
      \includegraphics[width=\textwidth]{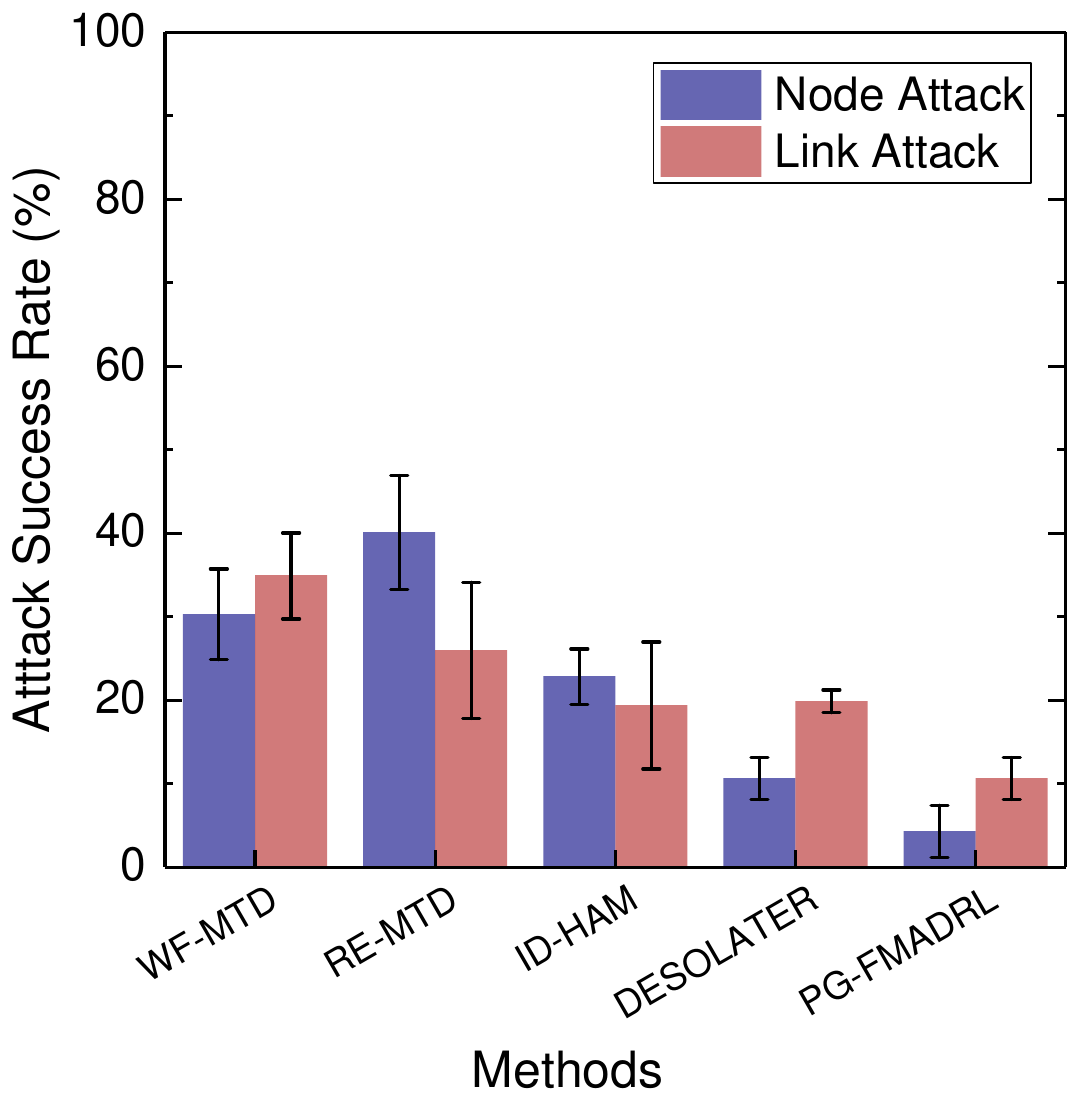}
    \end{minipage}
  }
  \hspace{-1mm}
  \subfigure[$N$ = 20]{
    \begin{minipage}[b]{0.18\textwidth}
      \includegraphics[width=\textwidth]{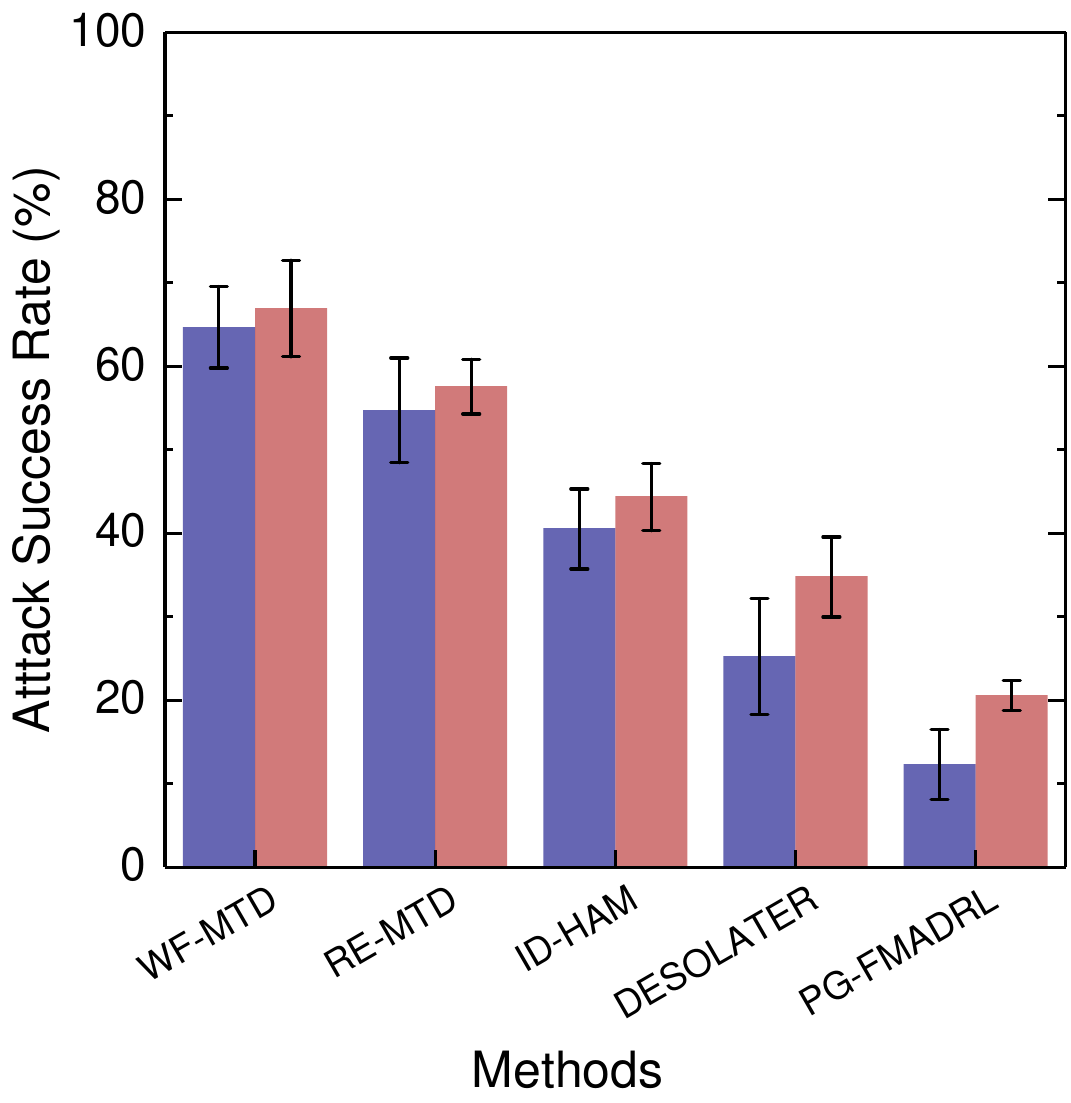}
    \end{minipage}
  }
  \hspace{-1mm}
  \subfigure[$N$ = 30]{
    \begin{minipage}[b]{0.18\textwidth}
      \includegraphics[width=\textwidth]{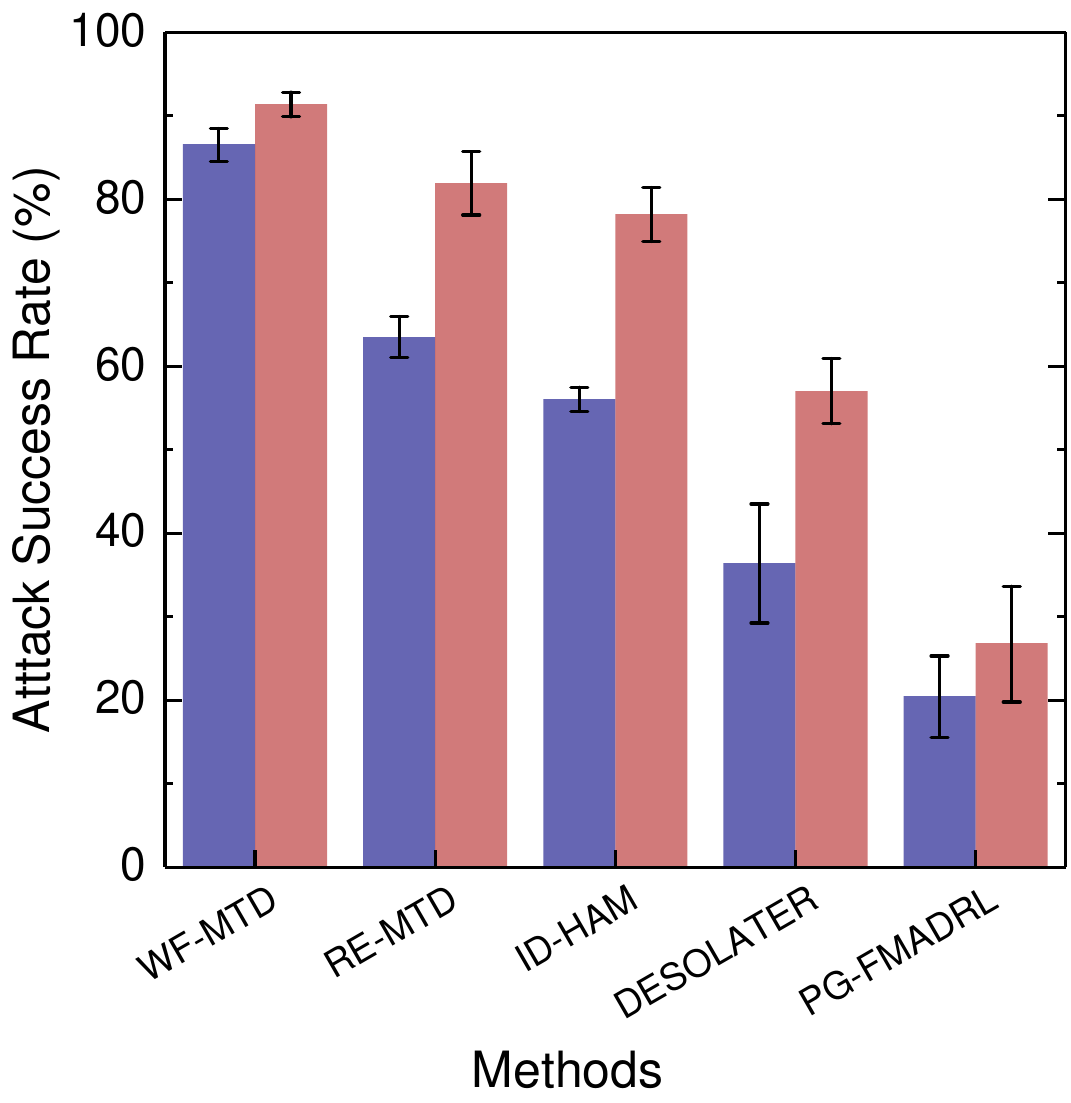}
    \end{minipage}
  }
  \hspace{-1mm}
  \subfigure[$N$ = 40]{
    \begin{minipage}[b]{0.18\textwidth}
      \includegraphics[width=\textwidth]{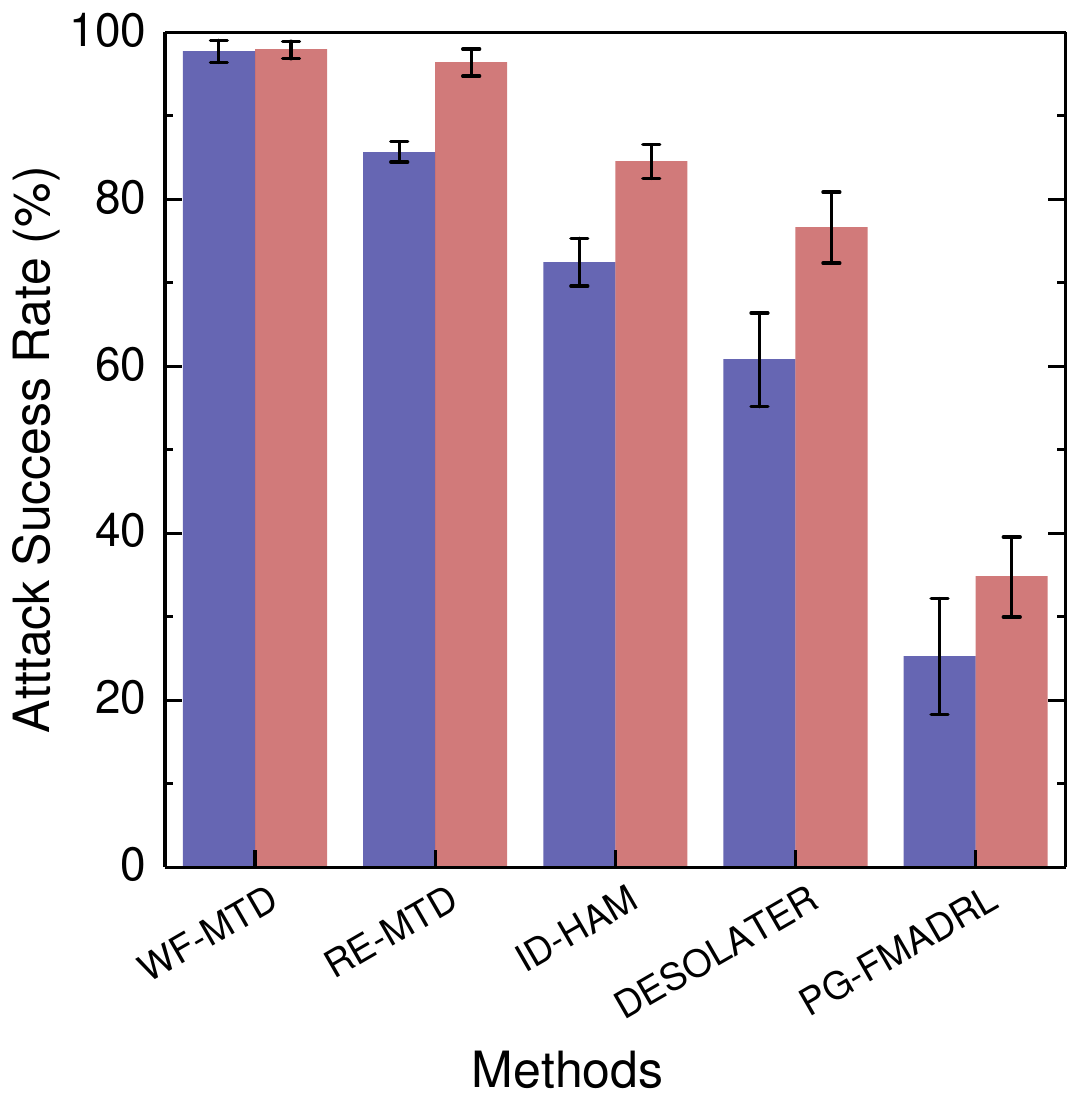}
    \end{minipage}
  }
  \hspace{-1mm}
  \subfigure[$N$ = 50]{
    \begin{minipage}[b]{0.18\textwidth}
      \includegraphics[width=\textwidth]{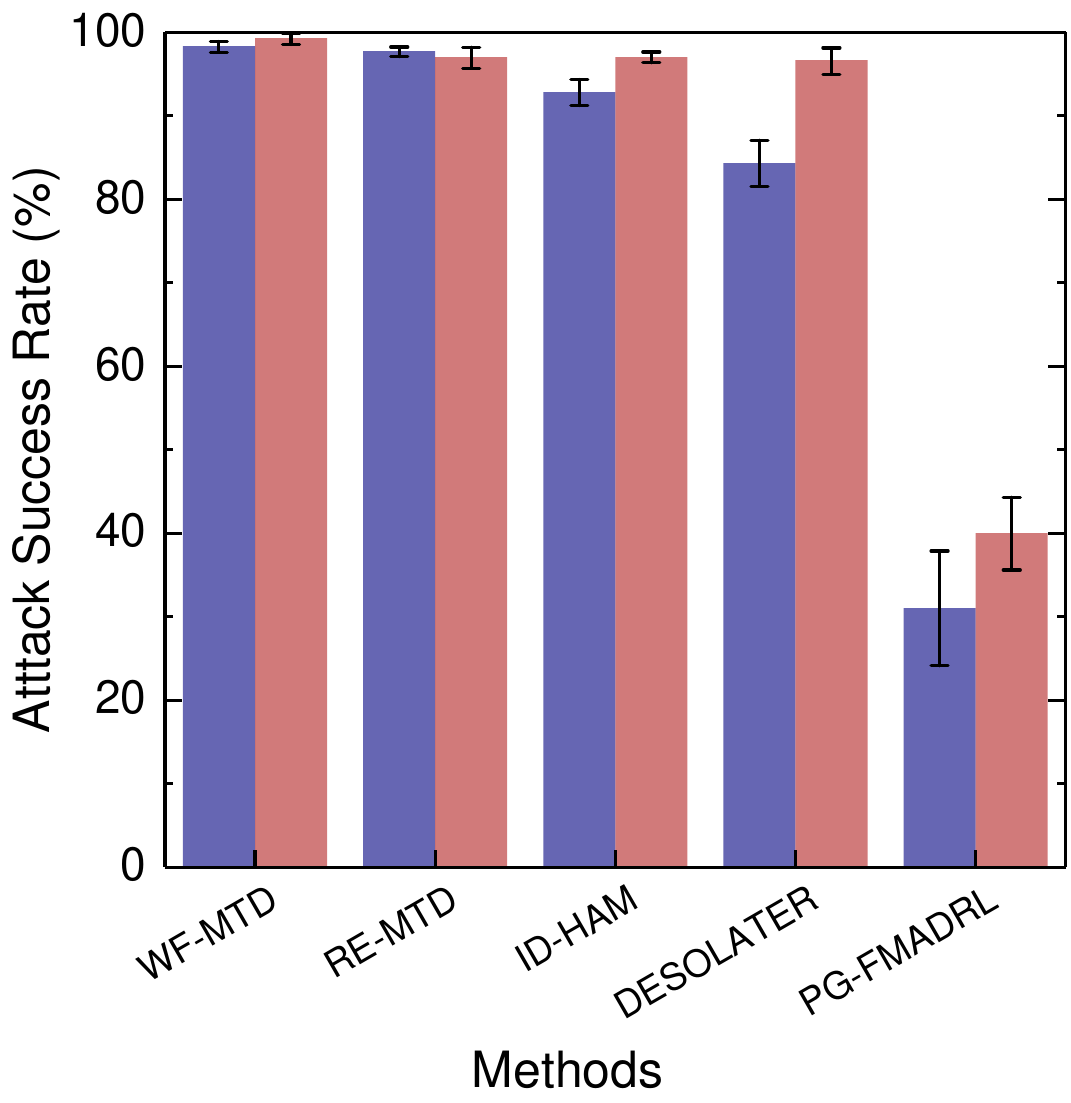}
    \end{minipage}
  }
  \caption{\textcolor{black}{The attack success rate for different defense methods under greedy node and link attacks. Each subplot corresponds to a specific swarm size ($N$).}}
  \label{Attack_success_rate}
\end{figure*}

When facing the more challenging greedy attacker, which dynamically adapts its strategy to maximize disruption, the advantage of PG-FMADRL becomes even more pronounced. While all baseline methods experience a sharp rise in cumulative cost, reflecting their struggle to efficiently counter adaptive threats, PG-FMADRL still maintains a much slower cost growth. This is attributed to its federated learning framework, which enables global agent collaboration and more intelligent, context-aware defense decisions. Notably, DESOLATER and ID-HAM, which also employ multi-agent DRL, perform better than WF-MTD and RE-MTD but still lag behind PG-FMADRL, especially in the link attack scenario. This further highlights the benefit of parameter aggregation and reward-weighted policy updates in our approach, which allow for both effective adaptation and cost control.

The lower energy consumption and defense cost achieved by PG-FMADRL are attributed to its intelligent and context-aware action selection. By leveraging policy gradient optimization, each UAV can precisely determine when and which MTD mechanism to trigger, thus minimizing unnecessary actions. This fine-grained control ensures resource efficiency while maintaining strong protection, outperforming baseline methods that rely on more frequent or less targeted defense operations.

\subsection{\textcolor{black}{Scalability Analysis}}
A critical aspect of any UAV swarm framework is its ability to scale effectively as the number of agents increases. To evaluate the scalability of our proposed method, we conducted additional experiments with varying swarm sizes, specifically for $N \in \{10, 20, 30, 40, 50\}$ UAVs. Specifically, we focused on the greedy strategy, and selected attack success rate as the key performance indicator, which measures the proportion of time the attacker successfully disrupts a target UAV or link.

The results are presented in Fig.~\ref{Attack_success_rate}, where each subfigure compares the five defense methods for a fixed swarm size. Two general phenomena are evident across all subplots. First, as the swarm size increases, the attack success rate for all methods tends to rise. This is an expected outcome, as larger swarms present more complex coordination challenges and a larger attack surface for the adversary. Second, the success rate of greedy link attacks is consistently slightly higher than that of greedy node attacks across most methods and scales, which aligns with our earlier findings (e.g., in Table~\ref{Mitigation_rate}) and suggests that disrupting communication links has a more cascading effect on the swarm's integrity.

Despite these challenges, the key finding is that our proposed PG-FMADRL method demonstrates significantly superior scalability when compared to SOTA methods. In every tested scenario, PG-FMADRL consistently achieves the lowest attack success rate for both attack types. For instance, in the most challenging scenario with $N=50$, the attack success rates for baseline methods soar to $\geqslant$ 84.30\% for node attacks and $\geqslant$ 96.58\% for link attacks. In contrast, our method successfully constrains these rates to 31.00\% and 39.96\%, respectively. This robust scalability is attributed to its federated learning architecture. While the performance of independent learning methods degrades more sharply as each agent grapples with an increasingly complex local problem, our reward-weighted aggregation mechanism ensures that effective defense policies are efficiently shared and adapted across the entire, growing network. This allows the swarm to maintain a robust, collective defense posture even at large scales, confirming that PG-FMADRL is highly scalable for deployment in large-scale low-altitude networks.

\section{Conclusion and Future Work}\label{section6}
In this paper, we proposed a novel FMADRL-driven MTD framework to proactively and adaptively mitigate DoS attacks in UAV swarm networks. By designing three lightweight and coordinated MTD mechanisms (e.g., leader switching, route mutation, and frequency hopping) and formulating the defense problem as a multi-agent POMDP, our approach enables each UAV to autonomously select defense actions based on local observations while benefiting from collaborative learning through reward-weighted federated aggregation. Extensive simulation results demonstrated that the proposed PG-FMADRL method significantly outperforms state-of-the-art baselines in terms of attack mitigation rate, recovery time, energy consumption, and defense cost, while maintaining robust mission continuity under various attack scenarios. By enhancing the security and robustness of UAV swarms, our work provides a foundational step toward the reliable and scalable deployment of low-altitude networks, supporting the vision of intelligent, resilient, and trustworthy low-altitude economic infrastructures.

Despite these promising results, our framework has limitations that open avenues for future research. A key limitation is that while the stochastic nature of our PG-FMADRL method inherently disrupts these adversaries, its resilience against a co-evolving, DRL-based adversary remains an open question. Another limitation is the reliance on a central aggregator which introduces a potential single point of failure. Therefore, future work should focus on enhancing our framework in two key directions. First, by evaluating it within a true adversarial learning context, potentially by modeling the interaction as a zero-sum game to develop more robust defense strategies. Second, by exploring more resilient aggregation models, such as decentralized or hierarchical federated learning, to eliminate the single point of failure.

\section*{Acknowledgments}
This work was supported in part by the National Natural Science Foundation of China under Grant No. 62202097 and Grant No. 62072100, in part by China Postdoctoral Science Foundation under Grant No. 2024T170143 and Grant No. 2022M710677, and in part by Jiangsu Funding Program for Excellent Postdoctoral Talent under Grant No. 2022ZB137.

\bibliographystyle{IEEEtran}
\bibliography{IEEEabrv,Reference}

\begin{thebibliography}{10}
\providecommand{\url}[1]{#1}
\csname url@samestyle\endcsname
\providecommand{\newblock}{\relax}
\providecommand{\bibinfo}[2]{#2}
\providecommand{\BIBentrySTDinterwordspacing}{\spaceskip=0pt\relax}
\providecommand{\BIBentryALTinterwordstretchfactor}{4}
\providecommand{\BIBentryALTinterwordspacing}{\spaceskip=\fontdimen2\font plus
\BIBentryALTinterwordstretchfactor\fontdimen3\font minus
  \fontdimen4\font\relax}
\providecommand{\BIBforeignlanguage}[2]{{%
\expandafter\ifx\csname l@#1\endcsname\relax
\typeout{** WARNING: IEEEtran.bst: No hyphenation pattern has been}%
\typeout{** loaded for the language `#1'. Using the pattern for}%
\typeout{** the default language instead.}%
\else
\language=\csname l@#1\endcsname
\fi
#2}}
\providecommand{\BIBdecl}{\relax}
\BIBdecl

\bibitem{zuo2022unmanned}
Z.~Zuo, C.~Liu, Q.-L. Han, and J.~Song, ``Unmanned aerial vehicles: Control
  methods and future challenges,'' \emph{IEEE/CAA Journal of Automatica
  Sinica}, vol.~9, no.~4, pp. 601--614, 2022.

\bibitem{xie2025disrupting}
C.~Xie, J.~He, S.~Guo, J.~Wang, S.~Zhang, T.~Zhang, and T.~Xiang, ``Disrupting
  vision-language model-driven navigation services via adversarial object
  fusion,'' \emph{arXiv preprint arXiv:2505.23266}, 2025.

\bibitem{sun2025aerial}
G.~Sun, J.~Xiao, J.~Li, J.~Wang, J.~Kang, D.~Niyato, and S.~Mao, ``Aerial
  reliable collaborative communications for terrestrial mobile users via
  evolutionary multi-objective deep reinforcement learning,'' \emph{IEEE
  Transactions on Mobile Computing}, vol.~24, no.~7, pp. 5731--5748, 2025.

\bibitem{javed2024state}
S.~Javed, A.~Hassan, R.~Ahmad, W.~Ahmed, R.~Ahmed, A.~Saadat, and M.~Guizani,
  ``State-of-the-art and future research challenges in uav swarms,'' \emph{IEEE
  Internet of Things Journal}, vol.~11, no.~11, pp. 19\,023--19\,045, 2024.

\bibitem{cao2024computational}
P.~Cao, L.~Lei, S.~Cai, G.~Shen, X.~Liu, X.~Wang, L.~Zhang, L.~Zhou, and
  M.~Guizani, ``Computational intelligence algorithms for uav swarm networking
  and collaboration: A comprehensive survey and future directions,'' \emph{IEEE
  Communications Surveys \& Tutorials}, 2024.

\bibitem{wang2024generative}
J.~Wang, H.~Du, D.~Niyato, J.~Kang, S.~Cui, X.~Shen, and P.~Zhang, ``Generative
  ai for integrated sensing and communication: Insights from the physical layer
  perspective,'' \emph{IEEE Wireless Communications}, vol.~31, no.~5, pp.
  246--255, 2024.

\bibitem{tsao2022survey}
K.-Y. Tsao, T.~Girdler, and V.~G. Vassilakis, ``A survey of cyber security
  threats and solutions for uav communications and flying ad-hoc networks,''
  \emph{Ad Hoc Networks}, vol. 133, p. 102894, 2022.

\bibitem{wang2025safeguarding}
J.~Wang, J.~He, G.~Sun, Z.~Xiong, D.~Niyato, S.~Mao, D.~I. Kim, and T.~Xiang,
  ``Safeguarding isac performance in low-altitude wireless networks under
  channel access attack,'' \emph{arXiv preprint arXiv:2508.15838}, 2025.

\bibitem{tang2024secure}
H.~Tang, Y.~Chen, and I.~Ali, ``Secure distributed model predictive control for
  heterogeneous uav-ugv formation under dos attacks,'' \emph{IEEE Transactions
  on Intelligent Vehicles}, 2024.

\bibitem{yang2025resilient}
H.~Yang, Z.~Yu, M.~Fu, and Y.~Zhang, ``Resilient consensus control for multiple
  uavs with input saturation under dos attacks,'' \emph{IEEE Transactions on
  Cybernetics}, vol.~55, no.~3, pp. 1159--1171, 2025.

\bibitem{shi2025neural}
S.~Shi, S.~Wu, and B.~Wei, ``Neural-network-based event-triggered formation
  tracking for nonlinear multi-uav systems with switching topologies under dos
  attacks,'' \emph{IEEE Transactions on Automation Science and Engineering},
  vol.~22, pp. 11\,656--11\,667, 2025.

\bibitem{wang2025generative}
J.~Wang, H.~Du, Y.~Liu, G.~Sun, D.~Niyato, S.~Mao, D.~I. Kim, and X.~Shen,
  ``Generative ai based secure wireless sensing for isac networks,'' \emph{IEEE
  Transactions on Information Forensics and Security}, vol.~20, pp. 5195--5210,
  2025.

\bibitem{wang2023survey}
Z.~Wang, Y.~Li, S.~Wu, Y.~Zhou, L.~Yang, Y.~Xu, T.~Zhang, and Q.~Pan, ``A
  survey on cybersecurity attacks and defenses for unmanned aerial systems,''
  \emph{Journal of Systems Architecture}, vol. 138, p. 102870, 2023.

\bibitem{ucctu2021suggested}
G.~U{\c{c}}tu, M.~Alkan, {\.I}.~A. Do{\u{g}}ru, and M.~D{\"o}rterler, ``A
  suggested testbed to evaluate multicast network and threat prevention
  performance of next generation firewalls,'' \emph{Future Generation Computer
  Systems}, vol. 124, pp. 56--67, 2021.

\bibitem{zhong2024survey}
M.~Zhong, M.~Lin, C.~Zhang, and Z.~Xu, ``A survey on graph neural networks for
  intrusion detection systems: methods, trends and challenges,''
  \emph{Computers \& Security}, p. 103821, 2024.

\bibitem{zhao2024two}
F.~Zhao, B.~Yang, C.~Li, C.~Zhang, L.~Zhu, and G.~Liang, ``Two-layer consensus
  based on master-slave consortium chain data sharing for internet of
  vehicles,'' \emph{arXiv preprint arXiv:2411.10680}, 2024.

\bibitem{tan2023survey}
J.~Tan, H.~Jin, H.~Zhang, Y.~Zhang, D.~Chang, X.~Liu, and H.~Zhang, ``A survey:
  When moving target defense meets game theory,'' \emph{Computer Science
  Review}, vol.~48, p. 100544, 2023.

\bibitem{zhang2025moving}
T.~Zhang, F.~Kong, D.~Deng, X.~Tang, X.~Wu, C.~Xu, L.~Zhu, J.~Liu, B.~Ai,
  Z.~Han \emph{et~al.}, ``Moving target defense meets artificial
  intelligence-driven network: A comprehensive survey,'' \emph{IEEE Internet of
  Things Journal}, vol.~12, no.~10, pp. 13\,384--13\,397, 2025.

\bibitem{zhou2025resource}
Y.~Zhou, G.~Cheng, Z.~Ouyang, and Z.~Chen, ``Resource-efficient low-rate ddos
  mitigation with moving target defense in edge clouds,'' \emph{IEEE
  Transactions on Network and Service Management}, vol.~22, no.~1, pp.
  168--186, 2025.

\bibitem{fu2023machine}
R.~Fu, X.~Ren, Y.~Li, Y.~Wu, H.~Sun, and M.~A. Al-Absi,
  ``Machine-learning-based uav-assisted agricultural information security
  architecture and intrusion detection,'' \emph{IEEE Internet of Things
  Journal}, vol.~10, no.~21, pp. 18\,589--18\,598, 2023.

\bibitem{hassler2024cyber}
S.~C. Hassler, U.~A. Mughal, and M.~Ismail, ``Cyber-physical intrusion
  detection system for unmanned aerial vehicles,'' \emph{IEEE Transactions on
  Intelligent Transportation Systems}, vol.~25, no.~6, pp. 6106--6117, 2024.

\bibitem{he2023cgan}
X.~He, Q.~Chen, L.~Tang, W.~Wang, and T.~Liu, ``Cgan-based collaborative
  intrusion detection for uav networks: A blockchain-empowered distributed
  federated learning approach,'' \emph{IEEE Internet of Things Journal},
  vol.~10, no.~1, pp. 120--132, 2023.

\bibitem{wu2024zero}
Y.~Wu, M.~Chen, and M.~Chadli, ``Zero-sum-game-based distributed fuzzy adaptive
  self-triggered control of swarm uavs under intermittent communication and dos
  attacks,'' \emph{IEEE Transactions on Fuzzy Systems}, vol.~32, no.~9, pp.
  5371--5384, 2024.

\bibitem{tan2025strategy}
J.~Tan, T.~Zheng, H.~Jin, Y.~Liu, H.~Zhang, and Z.~Tian, ``A strategy-making
  method for piot plc honeypoint defense against attacks based on the
  time-delay evolutionary game,'' \emph{IEEE Transactions on Information
  Forensics and Security}, 2025.

\bibitem{wu2022secure}
C.~Wu, W.~Yao, W.~Pan, G.~Sun, J.~Liu, and L.~Wu, ``Secure control for
  cyber-physical systems under malicious attacks,'' \emph{IEEE Transactions on
  Control of Network Systems}, vol.~9, no.~2, pp. 775--788, 2022.

\bibitem{sanoussi2023itc}
N.~Sanoussi, K.~Chetioui, G.~Orhanou, and S.~El~Hajji, ``Itc: Intrusion
  tolerant controller for multicontroller sdn architecture,'' \emph{Computers
  \& Security}, vol. 132, p. 103351, 2023.

\bibitem{ribeiro2023detecting}
M.~A. Ribeiro, M.~S.~P. Fonseca, and J.~de~Santi, ``Detecting and mitigating
  ddos attacks with moving target defense approach based on automated flow
  classification in sdn networks,'' \emph{Computers \& Security}, vol. 134, p.
  103462, 2023.

\bibitem{zhang2023moving}
T.~Zhang, C.~Xu, Y.~Lian, H.~Tian, J.~Kang, X.~Kuang, and D.~Niyato, ``When
  moving target defense meets attack prediction in digital twins: A
  convolutional and hierarchical reinforcement learning approach,'' \emph{IEEE
  Journal on Selected Areas in Communications}, vol.~41, no.~10, pp.
  3293--3305, 2023.

\bibitem{zhou2022toward}
Y.~Zhou, G.~Cheng, Y.~Zhao, Z.~Chen, and S.~Jiang, ``Toward proactive and
  efficient ddos mitigation in iiot systems: A moving target defense
  approach,'' \emph{IEEE Transactions on Industrial Informatics}, vol.~18,
  no.~4, pp. 2734--2744, 2022.

\bibitem{zhang2023mitigate}
T.~Zhang, C.~Xu, P.~Zou, H.~Tian, X.~Kuang, S.~Yang, L.~Zhong, and D.~Niyato,
  ``How to mitigate ddos intelligently in sd-iov: A moving target defense
  approach,'' \emph{IEEE Transactions on Industrial Informatics}, vol.~19,
  no.~1, pp. 1097--1106, 2023.

\bibitem{wang2021security}
L.~Wang, Y.~Chen, P.~Wang, and Z.~Yan, ``Security threats and countermeasures
  of unmanned aerial vehicle communications,'' \emph{IEEE Communications
  Standards Magazine}, vol.~5, no.~4, pp. 41--47, 2021.

\bibitem{gong2023resilient}
X.~Gong, M.~V. Basin, Z.~Feng, T.~Huang, and Y.~Cui, ``Resilient time-varying
  formation-tracking of multi-uav systems against composite attacks: A
  two-layered framework,'' \emph{IEEE/CAA Journal of Automatica Sinica},
  vol.~10, no.~4, pp. 969--984, 2023.

\bibitem{yu2023reinforcement}
Y.~Yu, W.~Yang, W.~Ding, and J.~Zhou, ``Reinforcement learning solution for
  cyber-physical systems security against replay attacks,'' \emph{IEEE
  Transactions on Information Forensics and Security}, vol.~18, pp. 2583--2595,
  2023.

\bibitem{boualouache2022federated}
A.~Boualouache and T.~Engel, ``Federated learning-based scheme for detecting
  passive mobile attackers in 5g vehicular edge computing,'' \emph{Annals of
  Telecommunications}, vol.~77, no.~3, pp. 201--220, 2022.

\bibitem{huang2022strategic}
M.~Huang, K.~F.~E. Tsang, Y.~Li, L.~Li, and L.~Shi, ``Strategic dos attack in
  continuous space for cyber-physical systems over wireless networks,''
  \emph{IEEE Transactions on Signal and Information Processing over Networks},
  vol.~8, pp. 421--432, 2022.

\bibitem{celdran2024rl}
A.~H. Celdr{\'a}n, P.~M.~S. S{\'a}nchez, J.~Von Der~Assen, T.~Schenk, G.~Bovet,
  G.~M. P{\'e}rez, and B.~Stiller, ``Rl and fingerprinting to select moving
  target defense mechanisms for zero-day attacks in iot,'' \emph{IEEE
  Transactions on Information Forensics and Security}, vol.~19, pp. 5520--5529,
  2024.

\bibitem{lei2024multi}
H.~Lei, D.~Meng, H.~Ran, K.-H. Park, G.~Pan, and M.-S. Alouini, ``Multi-uav
  trajectory design for fair and secure communication,'' \emph{IEEE
  Transactions on Cognitive Communications and Networking}, 2024.

\bibitem{chai2024system}
R.~Chai, B.~Wang, R.~Sun, X.~Jing, and Q.~Chen, ``System cost function
  optimization-based data scheduling and flight trajectory for multi-antenna
  uav-assisted communication and sensing integration systems,'' \emph{IEEE
  Transactions on Cognitive Communications and Networking}, pp. 1--1, 2024.

\bibitem{xu2025low}
J.~Xu, H.~Yao, R.~Zhang, T.~Mai, and M.~Guizani, ``Low latency and
  accuracy-guaranteed dnn inference for uav-assisted iot networks,'' \emph{IEEE
  Transactions on Cognitive Communications and Networking}, 2025.

\bibitem{tan2022wf}
J.~Tan, H.~Jin, H.~Hu, R.~Hu, H.~Zhang, and H.~Zhang, ``Wf-mtd: Evolutionary
  decision method for moving target defense based on wright-fisher process,''
  \emph{IEEE transactions on dependable and secure computing}, vol.~20, no.~6,
  pp. 4719--4732, 2022.

\bibitem{zhang2023how}
T.~Zhang, C.~Xu, J.~Shen, X.~Kuang, and L.~A. Grieco, ``How to disturb network
  reconnaissance: A moving target defense approach based on deep reinforcement
  learning,'' \emph{IEEE Transactions on Information Forensics and Security},
  vol.~18, pp. 5735--5748, 2023.

\bibitem{yoon2021desolater}
S.~Yoon, J.-H. Cho, D.~S. Kim, T.~J. Moore, F.~Free-Nelson, and H.~Lim,
  ``Desolater: Deep reinforcement learning-based resource allocation and moving
  target defense deployment framework,'' \emph{IEEE Access}, vol.~9, pp.
  70\,700--70\,714, 2021.

\bibitem{liu2017power}
Z.~Liu, R.~Sengupta, and A.~Kurzhanskiy, ``A power consumption model for
  multi-rotor small unmanned aircraft systems,'' in \emph{2017 international
  conference on unmanned aircraft systems (ICUAS)}.\hskip 1em plus 0.5em minus
  0.4em\relax IEEE, 2017, pp. 310--315.

\end{thebibliography}

\vspace{-25pt}
\begin{IEEEbiography}[{\includegraphics[width=1in,height=1.25in,clip,keepaspectratio]{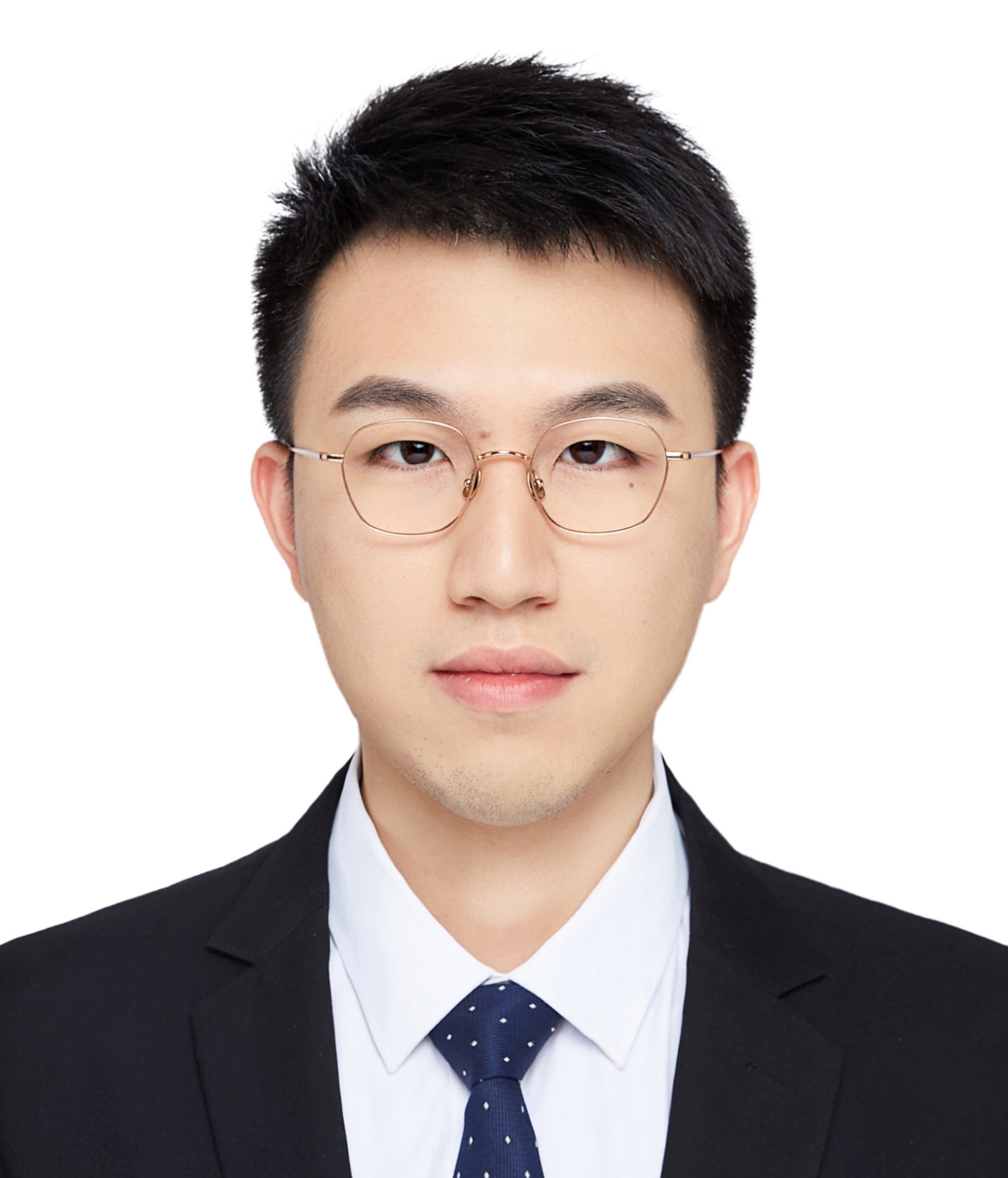}}]{Yuyang Zhou} received the Ph.D. degree in Cyberspace Security from Southeast University in 2021. He is currently working as a postdoc with the School of Cyber Science and Engineering, Southeast University. His major research interests include moving target defense, DDoS mitigation, and intrusion detection. He has published in some of the topmost journals and conferences like IEEE TIFS, IEEE TII, and ACM CCS, and is involved as reviewer and in technical program committees of several journals and conferences in the field.
\end{IEEEbiography}
\vskip-0.4in
\begin{IEEEbiography}[{\includegraphics[width=1in,height=1.25in,clip,keepaspectratio]{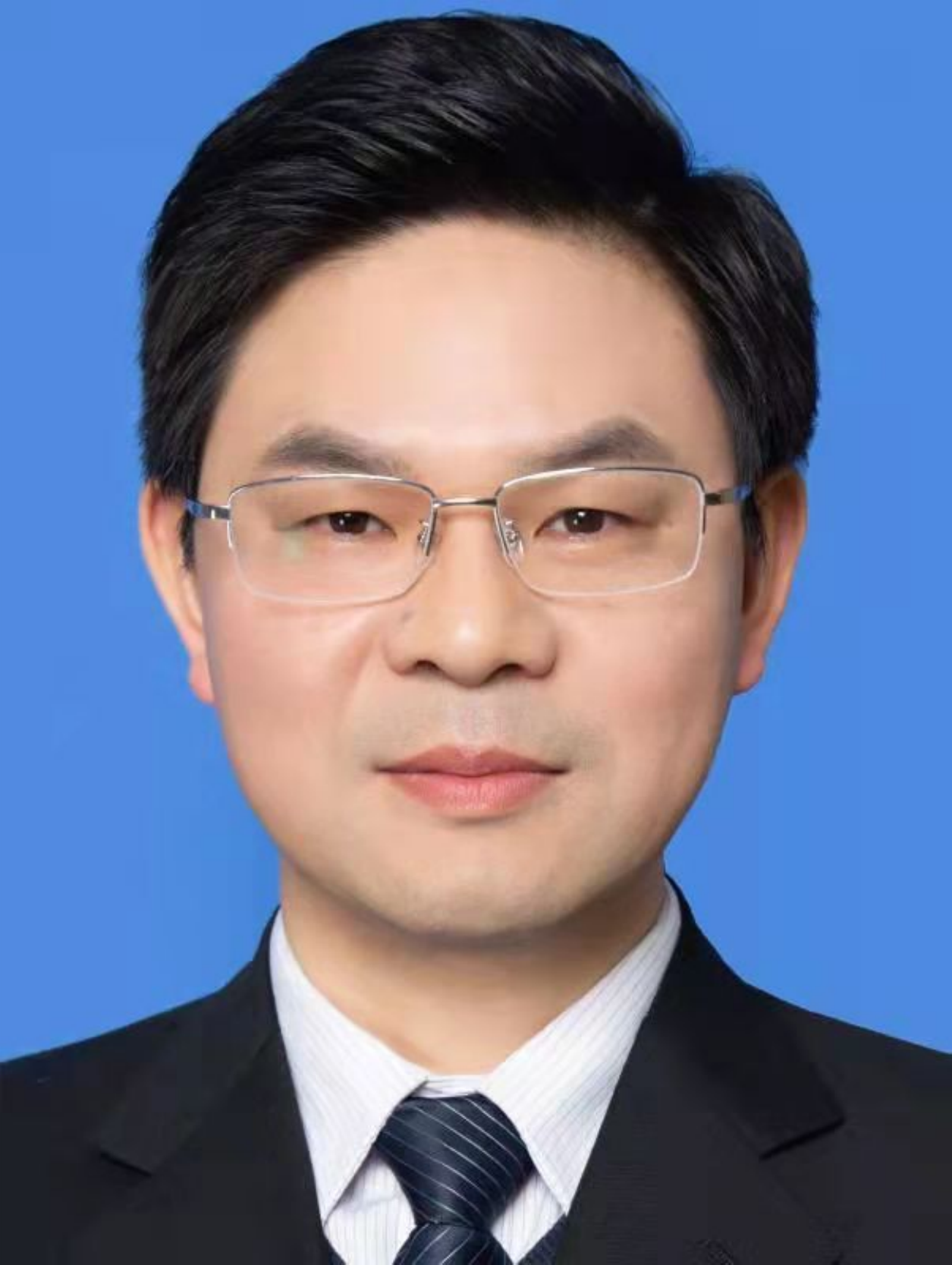}}]{Guang Cheng} received the Ph.D. degree in Computer Network from Southeast University in 2003. He is a Full Professor in the School of Cyber Science and Engineering, Southeast University, Nanjing, China. He has authored or coauthored seven monographs and more than 200 technical papers, including top journals and conferences like IEEE ToN, IEEE TIFS, IEEE TDSC, and IEEE INFOCOM. His research interests include network security, network measurement, and traffic behavior analysis. 
\end{IEEEbiography}
\vskip-0.4in
\begin{IEEEbiography}[{\includegraphics[width=1in,height=1.25in,clip,keepaspectratio]{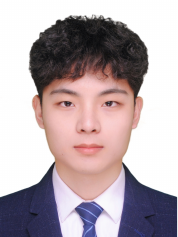}}]{Kang Du} received the B.S. degree in computer science and technology from Xidian University in 2023. He is currently pursuing the master's degree with the School of Cyber Science and Engineering, Southeast University. His major research interests include moving target defense, DDoS mitigation, and microservices security.
\end{IEEEbiography}
\vskip-0.4in
\begin{IEEEbiography}[{\includegraphics[width=1in,height=1.25in,clip,keepaspectratio]{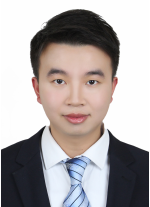}}]{Zihan Chen} obtained his Ph.D. degree in Cyber Security from Southeast University in 2023 and B.S. degree in Software Engineering from Central South University in 2017. He is currently working as a postdoc with the School of Cyber Science and Engineering at Southeast University. His major research interests include cyber security, encrypted traffic classification, encrypted traffic feature engineering, and deep learning.
\end{IEEEbiography}
\vskip-0.4in
\begin{IEEEbiography}[{\includegraphics[width=1in,height=1.25in,clip,keepaspectratio]{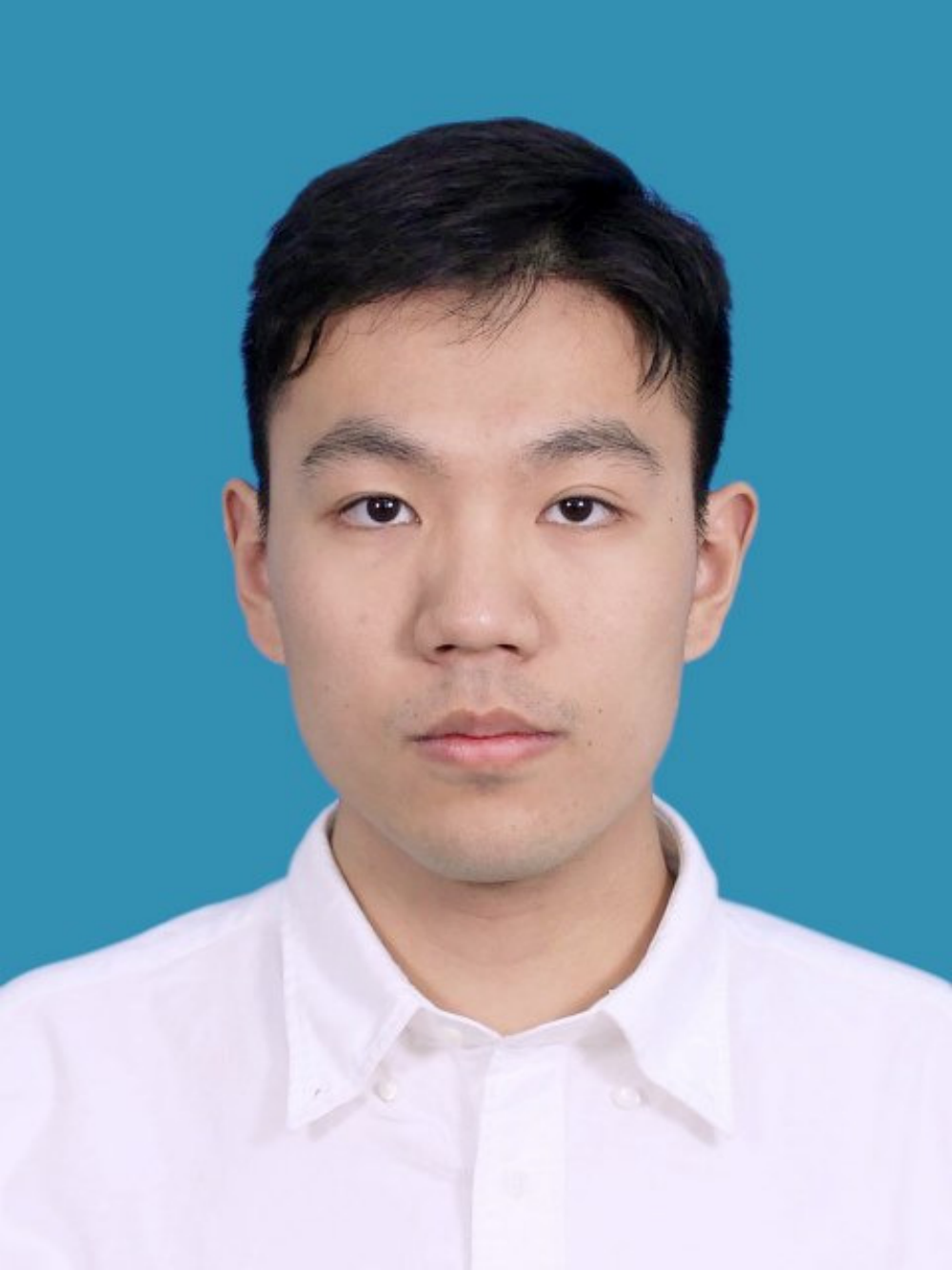}}]{Tian Qin} received the B.S. degree in information and computing sciences from Hohai University (HHU) in 2020. He is currently a Doctor candidate at the School of Cyber Science and Engineering, Southeast University, Nanjing, China. His current research interests include malicious traffic detection and federated learning.
\end{IEEEbiography}
\vskip-0.4in
\begin{IEEEbiography}[{\includegraphics[width=1in,height=1.25in,clip,keepaspectratio]{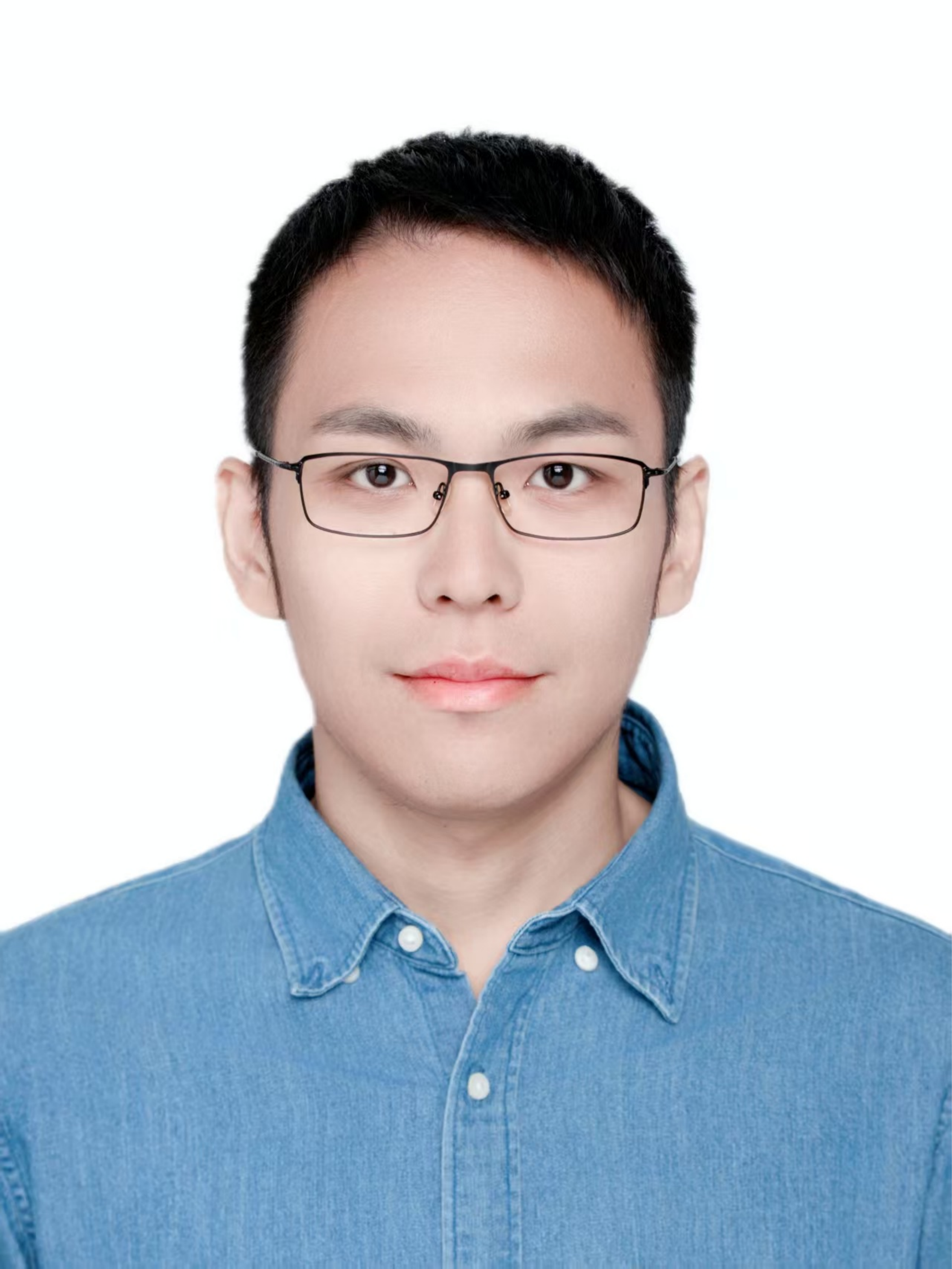}}]{Yuyu Zhao} received the B.S. degree in software engineering from the Nanjing University of Science and Technology in 2016 and the M.S. and Ph.D. degrees in cyber security from Southeast University, Nanjing, China, in 2019 and 2023, respectively. He is a currently a Lecturer with the School of Cyber Science and Engineering, Southeast University. His research interests include in-band telemetry, blockchain, and network processors.
\end{IEEEbiography}

\end{document}